\documentclass[twocolumn]{aastex61}%

\usepackage{graphicx}
\usepackage{booktabs}
\usepackage{multirow}
\usepackage{aas_macros}
\usepackage{hyperref} 
\usepackage{ulem}
\usepackage{natbib} 
\usepackage{amsmath}

\usepackage{capt-of}
\hypersetup{colorlinks,citecolor=blue,linkcolor=blue,urlcolor=blue}
\hypersetup{draft}

\newcommand{\revision}[1]{#1}
\newcommand{\revisiontwo}[1]{#1}

\shorttitle{RGZ: Knowledge Transfer Using PINK}
\shortauthors{Galvin et al.}

\begin{document}

\title{Radio Galaxy Zoo: Knowledge Transfer Using Rotationally Invariant Self-Organising Maps}

\author{T. J. Galvin}
\affiliation{CSIRO Astronomy and Space Science, PO Box 1130, Bentley WA 6102, Australia}
\affiliation{Western Sydney University, Penrith Campus, Locked Bag 1797, Penrith NSW 2751}

\author{M. Huynh}
\affiliation{CSIRO Astronomy and Space Science, PO Box 1130, Bentley WA 6102, Australia}
\affiliation{International Centre for Radio Astronomy Research (ICRAR), M468, The University of Western Australia, 35 Stirling Highway, Crawley, WA 6009, Australia}

\author{R. P. Norris}
\affiliation{Western Sydney University, Penrith Campus, Locked Bag 1797, Penrith NSW 2751}
\affiliation{CSIRO Astronomy and Space Science, PO Box 76, Epping, NSW 1710, Australia}

\author{X. R. Wang}
\affiliation{CSIRO Data61, Australia, Corner of Vimiera and Pembroke Roads, Marsfield NSW 2122, Australia}

\author{E. Hopkins}
\affiliation{Astroinformatics, HITS gGmbH, Schloss-Wolfsbrunnenweg 35, 69118 Heidelberg, Germany}

\author{O.\ I. Wong}
\affiliation{International Centre for Radio Astronomy Research (ICRAR), M468, The University of Western Australia, 35 Stirling Highway, Crawley, WA 6009, Australia}

\author{S. Shabala}
\affiliation{School of Natural Sciences, University of Tasmania, Private Bag 37, Hobart, Tasmania 7001, Australia}

\author{L. Rudnick}
\affiliation{Minnesota Institute for Astrophysics, University of Minnesota}

\author{M. J. Alger}
\affiliation{Research School of Astronomy and Astrophysics, The Australian National University, Canberra, ACT 2611, Australia}
\affiliation{Data61, CSIRO, Canberra, ACT 2601, Australia}

\author{K. L. Polsterer}
\affiliation{Astroinformatics, HITS gGmbH, Schloss-Wolfsbrunnenweg 35, 69118 Heidelberg, Germany}

\begin{abstract}
With the advent of large scale surveys the manual analysis and classification of individual radio source morphologies is rendered impossible as existing approaches do not scale.
The analysis of complex morphological features in the spatial domain is a particularly important task.
Here we discuss the challenges of transferring crowdsourced labels \revision{obtained from the Radio Galaxy Zoo project} and introduce a proper transfer mechanism via \revisiontwo{quantile random forest regression}.
By using parallelized rotation and flipping invariant Kohonen-maps, \revision{image cubes of Radio Galaxy Zoo selected galaxies formed from the FIRST radio continuum and WISE infrared all sky surveys} are first projected down to a two-dimensional embedding in an unsupervised way.
This embedding can be seen as a discretised space of shapes with the coordinates reflecting morphological features as expressed by the automatically derived prototypes.
\revision{We find that these prototypes have reconstructed physically meaningful processes across two channel images at radio and infrared wavelengths in an unsupervised manner. }
In the second step, images are compared with those prototypes to create a heat-map, which is the 
morphological fingerprint of each object and the basis for transferring the user generated labels. 
\revision{These heat-maps have reduced the feature space by a factor of 248 and are able to be used as the basis for subsequent ML methods. Using an ensemble of decision trees we achieve upwards of \revisiontwo{85.7\% and 80.7\% accuracy when predicting the number of components and peaks in an image, respectively, using these heat-maps. }}
We also question the currently used discrete classification schema and introduce a  continuous scale that better reflects the uncertainty in transition between two classes, caused by sensitivity and resolution limits.

\end{abstract}


\section{INTRODUCTION }
\label{sec:intro}
Radio astronomy is on the verge of a new age as the next generation of instruments nears  completion \citep{2017NatAs...1..671N...1..671N}.
These new instruments offer improvements in sensitivity, fractional bandwidth coverage and survey speed offering orders of magnitude improvement over conventional instruments, enabling us to unlock and explore a younger Universe. 

Associating a radio source with a single, intrinsic object across multi-wavelength domains is a difficult problem.
For example, different wavelengths can trace different physical emission mechanisms, which may not necessarily be localized in a single, compact region.
Radio lobes of Active Galactic Nuclei (AGN) may be separated by some distance from the super massive black hole accreting and ejecting matter, while only the host galaxy is seen at optical and infrared wavelengths. 
For such sources, it is important to correctly associate these physically separate components spanning different wavelength domains to extract the maximum level of scientific knowledge.
\revisiontwo{For instance, without multi-wavelength data the radio lobes of an AGN may be confused with two nearby, unrelated Star Forming Galaxies (SFG).}
A challenge is that the spatial resolution and sensitivity may be insufficient to separate them into distinct independent classes.

This problem is further exacerbated by the fact that different telescopes, and the data that they produce, have different characteristics and limitations.
For instance, optical and infrared surveys typically have higher resolution than radio surveys.
As a result, an infrared image may show many  objects in the vicinity of a single radio object.
Differing sensitivity limits may also influence the number of objects detected, and make identifying associated components harder, particularly if they are faint or missing in a subset of images.

Automated algorithms for such problems are in their infancy.
Although near-neighbour matching algorithms are generally robust for unresolved objects, the problem is more difficult to solve reliably for complex morphologies \citep{2018MNRAS.tmp.1258A}.
It is estimated that $10\,\%$ of the 70 million objects to be detected by \revisiontwo{Evolutionary Map of the Universe (EMU)} will be complex objects \citep{2011PASA...28..215N} requiring sophisticated methods of cross-identification.
Experts in the domain area (i.e. astronomers) will be unable to maintain pace to manually inspect every instance. 

With the advent of high performance computing platforms, commodity computing hardware is now capable of  solving the problem.
In particular, machine learning (ML) algorithms offer powerful avenues for both supervised and unsupervised data processing, classification and analysis, with applications ranging from photometric redshift estimation \citep{luken2018,norris2018}, star classification \citep{1995AJ....109.2401W}, optical transients \citep{2011BASI...39..387M} and simple/complex object discrimination \citep{2018arXiv180510718S,park2018,2018MNRAS.476..246L}.
Combining currently available large data-sets with the affordable computing resources provided by, e.g. graphics processing units (GPUs), opens novel data analysis techniques.

For image classification, considerable progress has been made by using convolution neural networks (CNN). 
\revision{These networks efficiently recognize hierarchical structures through a series of layered convolution functions after an initial training process.} 
Convolutional filters provide positional invariance to the feature location, which is invaluable when attempting to classify galaxies as they appear across the sky.
To deal with rotation invariance, spatial transformation layers that implement e.g. chirp z-transformations or data augmentation are common tools.

Several projects have successfully applied CNNs to galaxy classification problems. 
\citet{2017ApJS..230...20A} used CNNs to recognize classes of Fanaroff-Riley \citep[FR; ][]{1974MNRAS.167P..31F} radio galaxies and radio galaxies with bent tail morphologies.
They found success rates upwards of 95\% depending on the morphology presented, with bent-tailed radio galaxies being the most distinguishable.
\citet{2018arXiv180512008W} presents a CNN architecture that is capable of recognizing radio source morphologies, with an initial end-to-end classifier that is both fast ($<200$ milliseconds per image) and accurate ($>90\%$).
\revisiontwo{\citet{2018MNRAS.tmp.1258A} compared the performance of a simple CNN, random forests and linear regression in the task of classifying Radio Galaxy Zoo (RGZ) sources.}
\citet{2018MNRAS.476..246L} train a CNN on four classes of extended and compact sources,  achieving an overall classification accuracy of 94.8\% on the \revisiontwo{RGZ Data Release 1 (DR1)} data-set.

CNNs are an example of a supervised learning method, meaning that data-sets with known labels or features have to be provided for the training process to converge and become a successful predictor.
For certain problems, this can be a non-trivial requirement, as
the known data-set has to be sufficiently large and contain adequate sampling of the desired features or labels to be modelled.
Building such data-sets is often the most troubling task when attempting to utilize CNNs or similar supervised learning methods. 

Projects that use crowdsourcing methods to build these training data-sets for galaxy classification include
 Galaxy Zoo \citep{2008MNRAS.389.1179L} and \revisiontwo{RGZ} \citep{2015MNRAS.453.2326B}, both of which are members of the `Zooniverse' portal\footnote{\url{https://www.zooniverse.org/}}.
These projects provide an online web platform that allows volunteers to interact and label images.
Statistics are then built up of each source through a number of independent, non-experts classifications.
Although the classifications are performed by the general public, \revision{who are acting as citizen scientists and} may have no formal astronomy training, the consensus is generally comparable with expert classification. 

An alternative approach is to apply unsupervised ML methods which require no training set of  labels or features, but instead attempts to construct and optimize a function that is able to describe the \textit{structure} of the data.
Unsupervised clustering and dimensionality reduction methods are powerful tools to structure and explore large data-sets in such an unsupervised setting \citep{2016arXiv160105654G,2017ApJS..228...24T}.
Outlier detection and the search for the unexpected can also be considered unsupervised tasks \citep{2016arXiv161102829C, 2017PASA...34....7N}.
Self-Organising Maps \citep[SOM;][]{Kohonen1982} provide an unsupervised method to automatically derive a latent grid of discrete prototypes, where closeness in the projected space reflects closeness with respect to the used similarity measure.
\revisiontwo{SOMs have been used in the astronomical literature for a variety of tasks, including the classification of light curves \citep{2004MNRAS.353..369B}, clustering and analysis of gigahertz-peaked spectrum sources \citep{2008A&A...482..483T}, detecting structure within point data \citep{2011ApJ...727...48W} and object classification and photometric redshift estimation \citep{2012MNRAS.419.2633G}.  }

\revisiontwo{Applying the SOM method onto image data-sets requires special consideration}.
Even though the simple pixel-wise Euclidean distance between two images does not take spatial structures into account, it is already sufficient to order images by shape \revision{when rotational invariance is not an issue or when images have been aligned to a common orientation as part of some preprocessing stage.}
As the pixel values reflect locally measured intensities, \revision{pairs of objects} that have the same shape should have a distance close to zero.
With more and more pixels showing significantly different values, the represented shapes change together with the pixel-wise distance.

The Parallelized rotation and flipping INvariant Kohonen-maps \revisiontwo{\citep[PINK; ][]{polsterer2016}} software framework \revisiontwo{exploits GPU acceleration and} is designed to extend the basic SOM method to operate on image data where the simple rotation of a subject should not be a considered as part of the \textit{structure} of the data.
Similar sources are grouped, irrespective of their rotation and mirroring on the sky, which allows the resulting projection to be used to derive the distribution of shapes within the data-set and to recognize and separate unusual or rarely seen objects \citep{2016arXiv161102829C}.

In this paper, we assess the effectiveness of dimensionality reduction to transfer user generated labels to  unseen objects, using data from RGZ.

The paper outline is as follows.
In \S\ref{sec:SOM} we provide a brief description of the SOM algorithm and PINK, together with a description of the applied method to transfer labels.
We provide an outline of RGZ and the training data, the pre-processing steps and \revisiontwo{object} labels in \S\ref{sec:rgz}.
An overview of the  data experiments and the application of PINK is given in \S\ref{sec:apply_pink} with corresponding results being reported in \S\ref{sec:results}.
We finally provide points for discussion and conclusions in \S\ref{sec:discussion} and \S\ref{sec:conclusion}.

\section{Self-Organising Maps}
\label{sec:SOM}

A SOM is a commonly used algorithm to project high-dimensional data in a low dimensional space and thereby reflect similarity in the original space as distance in the \revisiontwo{SOM lattice}.
In the projected \revisiontwo{SOM lattice} space, ``closer'' refers to data being more similar and ``distant'' to be very different, without having a strict and formal connection between distance and similarity.
\revisiontwo{SOMs are neural networks that}, by being iteratively trained, learns how to arrange the pre-dominantly features in the input data.
Importantly, the training phase that produces the transformation to the lower dimensional space is unsupervised, requiring no input labels to accompany the data.
By specifying a similarity measure, the notion of distance in the high dimensional space can be used to, e.g. make a SOM aware of similar shape structures within spatially correlated image data.

The individual cells \revisiontwo{that make up the SOM lattice} are called neurons. \revisiontwo{Each neuron has its own set of weights, also called prototypes, and are constantly modified} during training to adapt for the incoming data in order to find a generalized representation.
Therefore the neurons are arranged in a low dimensional space, with 2D and 3D \revisiontwo{lattices} being the most common structures.

We provide a brief outline of the individual steps of the SOM algorithm, below:

\begin{enumerate}

\item {\it INITIALIZE}:
The prototypes are initialized with, e.g. noise, zeros, predefined patterns/structures, randomly drawn objects from the data set.
Alternatively, initialization could  be performed with a pre-trained set of prototypes.
\item {\it FIND BEST MATCH}:
An object is taken from the training data-set and its distance to all prototypes is calculated. The prototype, known as the Best Matching Unit (BMU), is the one with the minimum distance to the object.
Commonly used distance functions are the Euclidean distance, the Manhattan distance, and the Minkowski distances as their generalized counterpart.
\item {\it MODIFY MAP}:
Based on a \revisiontwo{neighborhood} function that is \revisiontwo{evaluated against neuron positions on the SOM lattice}, all prototypes are modified.
To realize an exponential decay \revisiontwo{between neighbouring neurons on the SOM lattice and the BMU, often a Gaussian function is selected. For instance, the Gaussian neighborhood function could be constructed as}

\begin{equation}
	r(n_1, n_2) = \frac{1}{\sqrt{2\pi\sigma^2}}e^{\frac{\left(n_1-n_2\right)^2}{2\sigma^2}}, 
\end{equation}

\noindent \revisiontwo{where $n_1$ and $n_2$ are the coordinates of two neurons on the SOM lattice, and $\sigma$ may be modified  between each iteration to focus more and more on a specific region of interest in prototype space to locally constrain the changes during training.}
In some cases a cyclic distance function that represents a continuous space that wraps around the edges could be considered.

The difference between each prototype and the current training object is calculated and used to modify the prototypes to be more like the current object.
Usually, \revisiontwo{neurons close to the BMU based on the neighbourhood function} are made more similar to the current training object, than \revisiontwo{neurons} further away. \revisiontwo{A basic weighting update scheme can be written as}

\revisiontwo{\begin{equation}
	w_{i}^\prime = w_{i} + \left(D - w_{i}\right) \times r\left(n_{i},BMU\right) \times \delta,
\end{equation}}

\noindent \revisiontwo{where $w_{i}$ and $w_{i}^\prime$ are the weights of the $i$-th neuron before and after performing the weighting update, $D$ is the currently select item from the training data-set, $r\left(n_{i},BMU\right)$ is the neighbourhood function evaluated using the $i$-th and current BMU neuron, and $\delta$ is the additional learning rate dampener that may also evolve across iterations. Its role is to further control the magnitude of the weighting updates between training iterations. }

\item {\it ITERATE}: Repeat steps 2 and 3 for $I$ number of iterations over all objects in your training data-set, where $I$ is sufficiently large to allow the SOM to converge.
\item {\it RECALL}: \revisiontwo{Once a sufficient number of training iterations have been performed to produce a stable SOM, map all objects in the data-set to the derived prototypes to determine the distances to the prototypes and find the region of best match.}

\end{enumerate}

The runtime complexity of this algorithm in a naive implementation is \revision{$O(I*N*M\mathrm{log}M)$}, depending on the number of iterations $I$, the number of training objects $N$ and the number of prototypes in the map $M$.
\revisiontwo{When using a GPU, the PINK software is able to operate in roughly $O(I*N*M)$ time when the number of GPU CUDA processing cores exceeds the number of generated images (see \S~\ref{sec:pink_ed_measure}). }
As this algorithm is not guaranteed to converge, it is important to determine the hyperparameters carefully.
The number of prototypes $M$ should be sufficiently large to represent the dominant structures in the training data.
Too many prototypes will result in a too large computation time, while too few prototypes will cause the map to shuffle without finally settling in a stable state.
If the learning rate is too high, the changes during training might be too abrupt, while too small values result in extremely long computations.
Too wide \revisiontwo{neighbourhood} functions, as used in step 3, change nearly all prototypes, while too narrow distance functions spatially decouples the individual \revisiontwo{neurons} on the \revisiontwo{SOM lattice} and therefore result in a simple clustering that is not reflecting gradual similarity between neighbouring prototypes.

\subsection{Rotation and Flipping Invariance}
\label{sec:pink_ed_measure}

PINK offers several distances to be used for the similarity function by offering pre-implemented distances as well as the ability to use functors to implement user-defined distances.
For our experiments, we used a modified Euclidean distance.
This is a simple metric, measuring the total distance between the pixel intensities of two images, following

\begin{equation}\label{eq:ed}
\begin{aligned}
\Delta(A,B) =& \\
\underset{\forall \phi \in \Phi}{minimize(\phi)}& \sqrt{\sum_{c=0}^{C}\sum_{x=0}^{X}\sum_{y=0}^{Y}\left(A_{c,x,y} - \phi (B_{c,x,y}) \right)^2},
\end{aligned}
\end{equation}

\noindent where $A$ and $B$ correspond to a particular neuron and image, $c$ is the corresponding channel, $x,y$ are the coordinates in the image plane and $\phi$ corresponds to an affine image transformation taken from a set of transformations $\Phi$.
In our case this set includes all possible rotations around the centre of the image as well as their mirrored counterparts.
To avoid empty patches at the corners that are caused by the rotation operation, PINK uses prototypes that are a factor of $\sqrt 2/2$ smaller than the input images.

As described in Equation~\ref{eq:ed}, the rotation invariance is introduced through a set of affine image transformations $\Phi$.
In principle the Euclidean distance is calculated for a set of possible rotations and the best matching angle is determined by finding the angle \revision{corresponding to the lowest distance}.
This job is done in parallel on a GPU hardware, providing the capability to do thousands of similar operations in parallel.
This brute-force comparison with all possible rotations ensures a rotation invariance with respect to the alignment morphological features.
Therefore the derived prototypes are rotation invariant representations of the data.

PINK does not include a damping functor or learning handicap term that evolves over time, to have a clear separation between the number crunching and the hyperparameter controlling part.
Although an early version did implement such a feature \citep{2015ASPC..495...81P}, the current version used throughout this study (version \texttt{0.23}) requires the user to vary these parameters across multiple invocations of the \revisiontwo{PINK program}. \revisiontwo{We list each invocation as a separate training stage in Table~\ref{tab:training}.}

\subsection{Comparison to Adaptive-Subspace SOM}

\revision{The adaptive-subspace SOM \citep{Kohonen1996} builds upon the basic SOM algorithm by attempting to learn invariance to affine image transformations throughout the training process. }
\revisiontwo{ This is done by mapping the training data into a collection of subspaces by applying various transforms, which may include randomly rotating an input image, shifting an image in some direction and zooming into a region of the image. }
\revisiontwo{ The BMU is located by exploring each subspace and all generated realisations of the transformed data in a `winner take all' manner. }
\revisiontwo{ Like the basic SOM algorithm, weight updates are shared among neighbouring neurons following some neighbourhood weighting function. }
\revisiontwo{ Transforming the data into these subspaces throughout training and mapping constructs a manifold that is able to project the high dimension input data onto a prototype lattice that is robust to affine transforms. }
\revisiontwo{ Computationally generating and exploring many of these realisations can be an expensive process. }

\revisiontwo{PINK adopts this approach but focuses on rotation and flipping invariance.}
\revisiontwo{ Within an astronomical context, these two types of transforms are the least significant when identifying objects and should largely be ignored.}
\revisiontwo{ Indeed, there are numerous object classification schemes which use scale or angular size as a distinguishing feature. }
\revisiontwo{ Preserving scale information throughout the training process will allow the constructed prototypes to represent more physically meaningful features that conform to existing astronomical morphologically classification schemes. }
\revisiontwo{Similarly, feature translation (i.e. location of a feature on an image) is largely a problem that is perhaps best addressed by source finding software. Broadly, these codes are tasked with locating objects in astronomical images to produce a catalogue of sources and their properties, including their central position. This task is not trivial as often domain specific knowledge has to be considered, including instrumental point spread functions, noise properties that may be correlated among adjacent pixels, coordinate system transforms and potentially varying wide field effects. }

\subsection{Heat-Maps as Morphological Fingerprints}
\label{subsection:heatmaps}

Once the SOM has been trained, then a heat-map can be produced for any \revisiontwo{object}.
A heat-map is a matrix of equal dimension to the \revisiontwo{SOM lattice}, whose values are the similarity measure of a source image (or image cube) to each \revisiontwo{prototype}.
In the case of PINK, this measure is the modified Euclidean distance (Equation~\ref{eq:ed}), and reflects the region of space that some image cube resides in the trained SOM out of the set of rotated and flipped images produced internally by PINK.
This heat-map is a single channel matrix (multi-channel image cubes and neurons are summed in this statistic).

The derived projection to the found prototypes can be seen as a projection to a space of shapes where individual regions represent characteristic morphologies.
Therefore the heat-maps represent a morphological fingerprint that characterizes the spatial structure of the image with respect to the automatically represented prototypes.

We convert  each modified Euclidean distance measure $\Delta(A,B_m)$ at the $m$-th position of the heat-map of a particular object to a likelihood $L_m$, \revision{by first normalising $\Delta(A,B_m)$ so its minimum is equal to one, then} using

\begin{equation}
\label{eq:ed-to-l-1}
L_{m}=\frac{\frac{\Delta(A,B_m)}{\Delta(A,B_m)^{\psi}}}  {\sum_{m=1}^{M} \frac{\Delta(A,B_m)}{\Delta(A,B_m)^{\psi}}},
\end{equation}

\noindent where $\psi$ is a stretching parameter which we nominally set to 10.
The purpose of the $\psi$ parameter is to introduce non-linearity in the transform between a simple Euclidean distance metric to a likelihood, where more emphasis is placed on the pixels with a smaller Euclidean distance.
It is important to note that by scaling with the sum over all individually stretched and transformed Euclidean distances the sum over all likelihoods is one.
Equation~\ref{eq:ed-to-l-1} is carried out as vector operations across all elements in the heat-map.

\subsection{Knowledge Transfer with Quantile Forest Regression}
\label{sec:rfr_outline}

A secondary method of attempting to classify \revisiontwo{objects} is to use their Euclidean distance or likelihood matrix as a whole, rather than selecting the single most likely neuron after building the label distribution across the SOM.
When projecting an input image onto a trained \revisiontwo{SOM lattice}, it is being placed into a lower dimension feature space. Rather than attempting to classify features in the image directly, the similarity measure produced by PINK can instead be thought of as a fingerprint from which a classification can be made. This lower dimensional space which is rotation invariant can be used as the basis for more generalised methods. 

A Random Forest Classifier \cite[RFC;][]{Breiman2001} is a supervised ML method which will construct a series of decision trees acting against an input set of training data to describe its corresponding labels.
Those decision trees can be seen as a segmentation of the feature space orthogonal to the dimension axis based on a specific information criterion.
To improve classification and control over-fitting, the input training data can be sub-divided to train a collection of individual decision trees, whose predictions are then collected and averaged.
Instead of using the mean prediction, the evaluation of the individual results is helpful to understand the distribution of the individual predictions further than the mean allows \citep{Meinshausen:2006:QRF:1248547.1248582}.
Therefore we built a quantile regression forest based on the standard \texttt{scikit-learn}\footnote{\url{http://scikit-learn.org/stable/index.html}} \citep{scikit-learn} \texttt{RandomForestRegressor} class by inspecting the predictions of all individual ensemble members, separately. 
\revision{While training the \texttt{RandomForestRegressor} we utilized a stratified $k$-fold cross validation strategy \citep{Mosteller:68,geisser_predictive_1975}. This approach randomly segments the data into $k$ number of sets, and across $k$ repetitions each segmented group is selected once for testing with the the remaining $k$-1 being used as training data. During each round the training segments of data were crafted to have an equal balance of class labels. For this work we use 5 folds.  }

The \texttt{RandomForestRegressor} was configured to construct an ensemble of 128 discrete decision trees individually constructed against bootstrapped sub-samples generated from the training data-set. 
\revision{We empirically selected the number of discrete trees as a compromise between accuracy of the \texttt{RandomForestRegressor} while being able to efficiently leverage all available CPU cores through parallelisation.  Other hyperparameters of the \texttt{RandomForestRegressor} were kept as their default values\footnote{Using \texttt{scikit-learn} version \texttt{0.19.1}}. }
Results from these individual trees were \revision{averaged} together to make a prediction.
Input features were the 2 dimensional likelihood matrices flattened to a one dimensional vector, which for a \revisiontwo{SOM lattice} of $15\times15$ neurons, constituted $M=225$ features.

The quantile regression forest using the individual cells of the heat-maps as input features can be used to transfer user generated labels to yet unseen objects. \revision{In the context of this work we use the quantile regression forest to predict the corresponding RGZ class labels of \revisiontwo{object} images that we describe in the following section. }

\section{Data from Radio GalaxyZoo}
\label{sec:rgz}

RGZ asks members to classify objects with complex radio morphologies across multiple wavelengths.
\revisiontwo{Upon participation, RGZ users become `citizen scientists' - members of the general public who undertake work in collaboration with professional scientists.}
Utilizing a web based interactive front end, the \revisiontwo{citizen scientists} are presented with a collection of multi-wavelength images and asked to classify various components and properties.
The idea is to generate a sample of answers from multiple \revisiontwo{responses} to build a consensus of what are the true radio and IR components of a single complex source.
Although individual \revisiontwo{citizen scientist participants} may not have domain expertise, the collective answers tend to be consistent with answers provided by experts in the field \citep{2015MNRAS.453.2326B}. 

Since its initial public launch, the service provides in excess of 170,000 radio source components for the \revisiontwo{citizen scientists} to classify using a collection of publicly available astronomy data-sets, which we describe below. 
\subsection{The Data}

For this experiment we obtained an internal pre-release copy of the RGZ Data Release 1 \citep{wong2018+} catalogue, which provides training labels produced by the citizen scientists of 103,930 radio components.
We summarize the data they used and their procedure for classification below roughly following \citet{2015MNRAS.453.2326B}.

The primary sample that makes up the RGZ database is sourced from the Very Large Array (VLA) 1.4\,GHz Faint Images of the Radio-Sky at Twenty centimeters \cite[FIRST; ][]{1994ASPC...61..165B}.
This program  covers roughly 10,000 square degrees of sky at a resolution of $5''$ down to a 1$\sigma$ r.m.s noise level of 150\,$\mu$Jy/beam. In all, FIRST detected approximately 947,000 discrete objects. 

Complementing this radio survey, RGZ used the Wide-field Infrared Survey Explorer (WISE) all sky program \citep{2010AJ....140.1868W}.
With four wavelength bands corresponding to 3.4, 4.6, 12 and 22\,$\mu$m (labelled as W1, W2, W3 and W4) reaching 5$\sigma$ point source sensitivities of 0.08, 0.11, 1. and 6.0\,mJy, respectively, the survey is a powerful tool to study the stellar and interstellar processes in galaxies. 

RGZ include in their data-set sources from the FIRST survey that satisfy two simple criteria: (1) the source has a signal to noise ratio in excess of 10, and (2) the source appears resolved. This second criterion excludes simple, compact radio sources, leaving complex sources with difficult-to-match morphologies.

\citet{2015MNRAS.453.2326B} define a source as being resolved if it satisfies

\begin{equation}
\frac{S_{\mathrm{peak}}}{S_{\mathrm{int}}} < 1.0 - \left(\frac{0.1}{\mathrm{log}\left(S_{\mathrm{peak}}\right)}\right),
\end{equation}

\noindent where $S_{\mathrm{peak}}$ is the flux density in units of mJy/beam and $S_{\mathrm{int}}$ is the total integrated flux of a source respectively.
Their final data set used for training the map consists of about 100,000 sources. 

For each object in DR1, we downloaded the corresponding Flexible Image Transport System \citep[FITS; ][]{1981A&AS...44..363W} images from the FIRST\footnote{\url{https://third.ucllnl.org/cgi-bin/firstcutout}} and WISE\footnote{\url{https://irsa.ipac.caltech.edu/ibe/docs/wise/allsky/4band_p3am_cdd/}} postage stamp services.

In total, we had access to images from the FIRST and WISE surveys of 103,930 objects with corresponding labels from RGZ DR1.
These labels presented in subsequent tables and figures encode the number of radio ``components'' ($N_C$) and ``peaks'' ($N_P$) as `$N_C\_N_P$'. Although the number of components is derived from the RGZ participants, the number of peaks is obtained as a product from the RGZ data processing pipeline.
The term ``component'' refers to discrete individual radio source components identified above a 4-sigma \revisiontwo{pixel intensity} threshold, and ``peak'' refers to the number of peaks within the set of components \citep{2015MNRAS.453.2326B}.
An object that appears to be a point-source may be classified as `1\_1', as it only has a single component with a single distinguishing peak.
An AGN whose jets have a small angular separation could be classified as `1\_2', as the single component of contiguous pixels would have \revisiontwo{multiple peaks}.
A more complex AGN with a distinction region of separation between its radio lobes may be classified as `2\_2', as it exhibits two individual components, each with one peak. 

\begin{figure}[h!]
  \includegraphics[width=\linewidth]{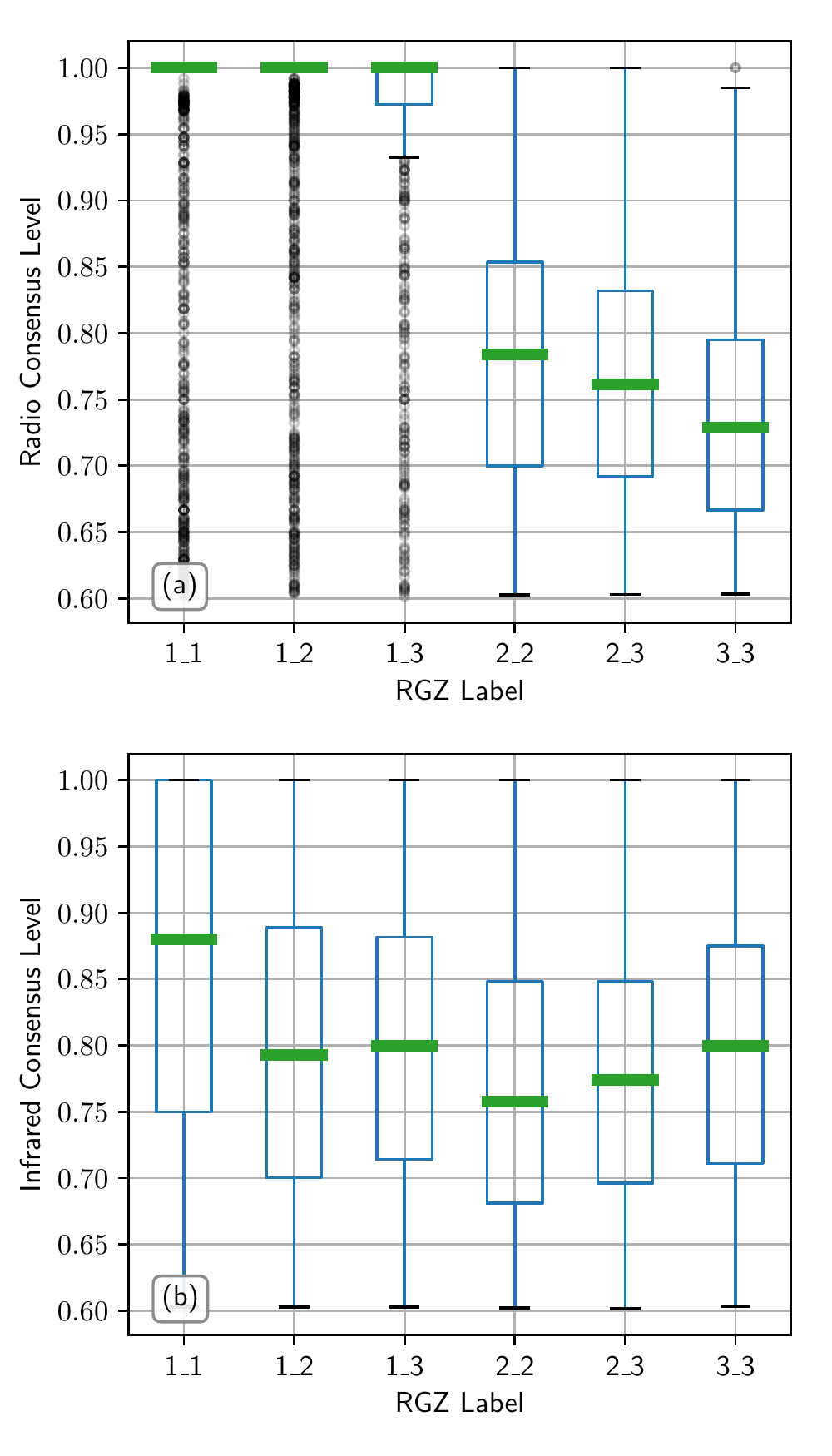}
  \caption{Distribution plots of the radio \textit{(a)} and infrared \textit{(b)} consensus levels of objects selected for the training and testing of label distribution across the SOM of six RGZ defined classes.
Each box spans from the lower to upper quartile, with the solid horizontal line representing the median value.
For classes whose interquartile range is zero as the majority of their CLs were one, only the median horizontal line is shown.
Whiskers from the boxes show the inter-quartile range extended by 150\%. Data points outside the whiskers are shown as circles. \label{fig:consensus_level} }

\end{figure}

Accompanying each object from RGZ DR1 is a consensus level (CL) which indicates how consistent a classification was.
It is defined by \citet{2015MNRAS.453.2326B} as $N_{\mathrm{consensus}}/N_{\mathrm{all}}$, where $N_{\mathrm{consensus}}$ is the number of volunteers who agree on the arrangement of the radio components, and $N_{\mathrm{all}}$ is the total number of classifications of an \revisiontwo{object}.
A CL closer to one indicates a more reliable classification, in the sense that a larger number of participants agreed.
For the label transfer experiments using a random forest regressor, objects with a high consensus level exceeding a value of $0.6$ have been chosen. A secondary selection was made so that labels matched the same set used by \citet{2018arXiv180512008W}. 
\revision{Applying this initial criteria produced a sample of about $50,000$ \revisiontwo{objects} with a highly imbalanced distribution of labels, where the more complex `1\_3', `2\_3' and `3\_3' classes having fewer then $800$ \revisiontwo{objects} in each. To better balance the class labels we randomly selected $2,000$ \revisiontwo{objects} in each, with repetition allowed. Duplicate \revisiontwo{object} labels were subsequently dropped.}
The distribution of the consensus levels across the individual classes is shown in Figure~\ref{fig:consensus_level} while numbers of those $7,464$ objects with respect to the provided labels are given in Table~\ref{tab:label_counts}. This subset of \revisiontwo{objects} are used for the random forest regression experiment. We note that more complex labels generally have a lower radio CL. Naturally, increased complexity and ambiguity among an \revisiontwo{object} image can lead to more disagreement among the participants. 

\begin{table}
  \centering
  \begin{tabular}{rrrrr}
  \hline
  Label & Components & Peaks & Number & Fraction\\
  \hline
  1\_1 & 1 & 1 & 1,947 & 26.1\% \\
  1\_2 & 1 & 2 & 1,786 & 23.9\% \\
  1\_3 & 1 & 3 & 775   & 10.4\% \\
  2\_2 & 2 & 2 & 1,585 & 21.2\% \\
  2\_3 & 2 & 3 & 631   & 8.5\% \\
  3\_3 & 3 & 3 & 740   & 9.9\% \\
  \hline
  Total & & & 7,464 & 100.0\%\\
  \hline
  \end{tabular}
  \caption{An overview of the types of labels and their counts which are used to build the random forest classifier. \label{tab:label_counts}} 
  \end{table}
  
No attempt is made to isolate or filter out images with potentially more than one discrete \revisiontwo{object} in the field. PINK will only learn consistent features. If by chance there are additional sources within the field secondary to the centered subject, these inconsistent features would be treated similar to noise. Throughout training these secondary sources, as they are inconsistent in terms of their proximity to the centered \revisiontwo{object}, would be filtered out in a manner similar to noise. 

\subsection{Data Pre-processing and Preparation}
\label{sec:preprocessing}

\begin{figure*}[ht]
  \includegraphics[width=\linewidth]{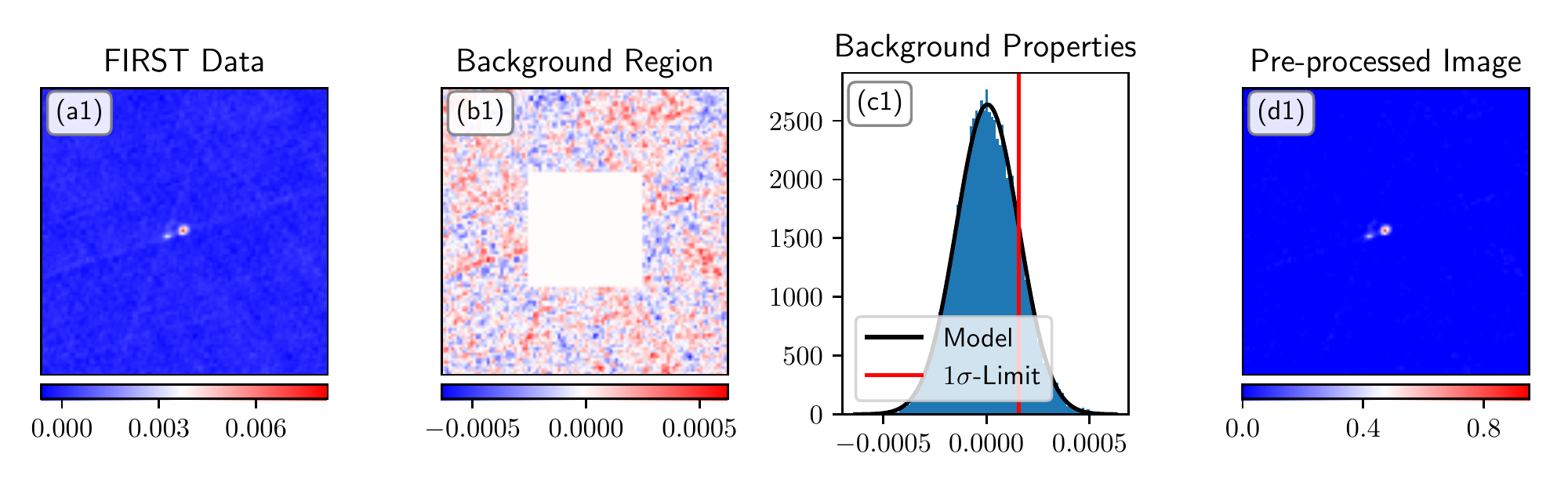}
  \includegraphics[width=\linewidth, trim=0cm 0cm 0cm 0cm, clip=true]{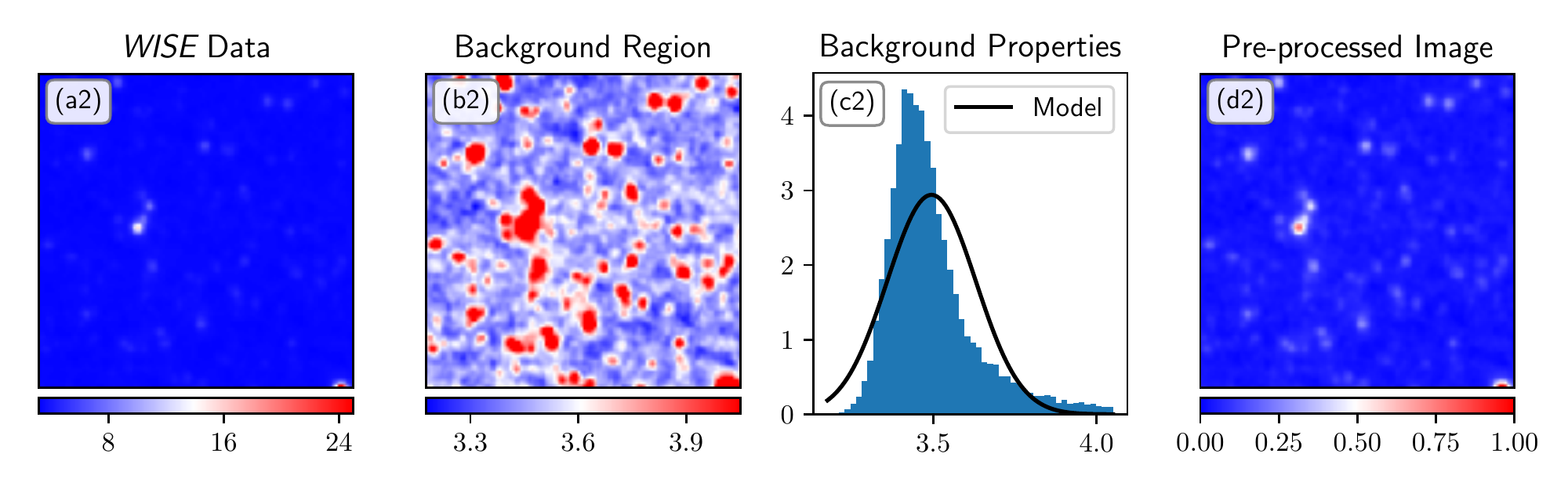}
  \caption{The pre-processing stages applied to the FIRST (top row) and \textsc{WISE} W1 band data (bottom row) data-sets  for \texttt{FIRST\,J140118.8+061210}. Column (a) shows the initial images from the FITS images.
In panels (b1) and (b2) we highlight the masking region used to obtain an estimate of the background noise statistics of the FIRST radio continuum image of each source.
The overlaid Gaussian in panel (c1) shows the model we construct to replace empty and missing pixel values.
The final FIRST pre-processed image, shown in (d1), places the pixels between the red vertical line in panel (c1) to the maximum pixel value onto a zero to one intensity scale.
Empty pixels in the \textsc{WISE} input image were replaced using the overlaid Gaussian model shown in panel (c2), based on all pixel intensities of the data presented in panel (b2). To better emphasize the noise characteristics we have applied a stretch to hide the brightest pixels of panel (a2). 
The \textsc{WISE} data were then placed onto a logged scale before being normalised onto a zero to one intensity scale, shown as the final pre-processed image in panel (d2).
We show the pixel intensity range under each appropriate panels as a accompanying colorbar.
Pixel values in the original FIRST and \textsc{WISE} images are Jy/beam and Digital Numbers (DN) respectively. \label{fig:preprocessing}}
\end{figure*}

The initial pre-processing step applied to all WISE W1 band FITS images was to reproject them onto the same pixel grid of the corresponding FIRST radio continuum image using the \texttt{astropy} \citep{2013A&A...558A..33A,2018arXiv180102634T} affiliated \texttt{python} based \texttt{reproject}\footnote{\url{https://reproject.readthedocs.io/en/stable/}} module.
PINK does not use the World Coordinate System \citep[WCS; ][]{2002A&A...395.1061G} and operates on binary data matrices.
Although this reprojection stage is not necessary for PINK to function, the convenience of having aligned features to the human observer across multi-wavelengths when preparing the data and inspecting the learnt features in the trained SOM was worth the small processing overhead, particularly when crafting image cubes whose channels contain images of different wavelength.
We applied distinct sets of pre-processing stages separately to the FIRST and WISE images, which we implement in the \texttt{ImagesToBin.py} script and describe below.

\subsubsection{FIRST}

The initial step applied to all FIRST images was searching for pixels whose values were designated as `Not a Number' (\texttt{NaN}).
These values represent pixels masked out of the imaging process, and were common for objects located near the edge of a FIRST mosaic region.
These values were removed by first characterizing the mean and standard deviation of valid pixels around the outer edge region and the image, and then randomly drawing a value assuming a normal distribution with the derived quantities to replace them with.
This ensures that the replacement of missing values does not introduce any morphological features.

Noise among the pixels of the FIRST radio continuum images is correlated due to the convolution of the Very Large Array (VLA) point spread function, even after applying the iterative \texttt{clean} algorithm \citep{1974A&AS...15..417H} and its more modern derivatives.
Applying ML methods to such data that \revisiontwo{exhibits} distinct structure in the background has to be done with care, as components or properties of the noise may be learnt as distinguishing features. 

After correcting the background fluxes for bias by subtracting the mean background pixel value from the image, all values below a one \revisiontwo{standard deviation} threshold are considered noise.
Therefore all values are shifted so that the one \revisiontwo{standard deviation} threshold is now the new zero point of the image and all negative values are clipped.
This is done to prevent the background from being considered in Equation~\ref{eq:ed} as a spatial characteristic.
Afterwards a scaling is applied to place all images onto a consistent intensity scaling, making the data intensity-invariant with respect to the applied distance function.

\subsubsection{WISE}

Unlike the FIRST data, the noise characteristics of WISE images are not correlated among nearby pixels as it is based on an infrared array.
The fact that there is no underlying structure in the noise means that PINK throughout its training process should not find features other than consistent source morphologies, aside from calibration issues.
No \revisiontwo{pixel} threshold clipping was applied to the WISE image data as it could in fact mask out real, faint features.

We replaced missing pixel values by sampling from a normal distribution whose mean and standard deviation were derived from the entire field of the image.
Although this included real emission, and is not a true characterisation of the image noise properties, the number of pixels replaced was minimal.
Image data were then placed on a logarithmic scale, and a min-max normalisation was then performed following
\begin{equation}
    I_{normalized} = \frac{I - \mathrm{min}\left(I\right)}{\mathrm{max}\left(I\right) - \mathrm{min}\left(I\right)},
    \end{equation}
    
\noindent where $I$ is the image data to be normalized.

\subsubsection{Image Cubes}

We created a two channel image cube using the pre-processed FIRST and WISE data compatible with the PINK binary format.
When crafting these final image cubes we applied a 95\% and 5\% weighting factor to the FIRST and WISE images, respectively.
This was done to encourage PINK to first focus on identifying the broad radio morphologies before isolating the WISE features.

We show these  steps of the pre-processing in Figure~\ref{fig:preprocessing} for \texttt{FIRST\,J140118.8+061210}.

\section{Experiments}
\label{sec:apply_pink}

\subsection{Training the SOM}
\label{sec:training_pink}

\revisiontwo{This work builds upon the maps presented by \citet{polsterer2016} by running } PINK  jointly on the FIRST and WISE data-sets, creating a single two channel map \revisiontwo{with radio and IR features}.
PINK itself has a number of user defined parameters that influence the algorithm and final convergence.

\revision{The SOM was initialised with random noise with a preferred direction. }
\revisiontwo{We used bilinear interpolation when generating rotated images. }
\revisiontwo{No periodic boundary conditions (i.e. edge wrapping neighbourhood function) were used.}
Training of the SOM for each experiment was carried out across five steps, each targeting and refining different feature scales across the surface of the SOM.
\revisiontwo{The neighbourhood function (which includes the learning rate) and desired number of rotations were set using the \texttt{--dist-func} and \texttt{--numrot} PINK arguments. Specific values used across different training stages are listed in Table~\ref{tab:training}.} 
\revisiontwo{A single iteration refers to using each item in the training data-set once when constructing the SOM. }

The goal of these consecutive stages is to establish the large scale structure and broad layout of source morphologies across the SOM surface, distinguish subsets of object types among the broad regions, and fine tune the neurons and their surrounding features.
\revisiontwo{A fine level of rotational increments at the earliest training stages is not needed, where only the broad structure of the SOM is established. The values presented in Table~\ref{tab:training} were empirically selected as a compromise between training time and accuracy. At the earliest training stages 48 interpolated images corresponds to $\sim8^\circ$ increments. The neuron dimensions are a factor of $\sqrt{2}/2$ smaller then then the original $5'\times5'$ input images. Assuming the worst case where a feature is on the border of the prototype, $8^\circ$ of rotation corresponds to a shift of roughly 20$\arcsec$. With the FIRST pixel size being $\sim1.8''$, this represents a potential misalignment of $11/2=6$\, pixels between a rotated image and an assumed prototype, in the worst case.} 

\begin{table}
  \centering
  \begin{tabular}{rrrrr}
  \hline
  Training & Sigma & Learning & Rotations & Iterations \\
  Stage & & Rate & & \\
  \hline
  1 & 1.5 & 0.10 & 48 & 2 \\
  2 & 1.0 & 0.05 & 92 & 5 \\
  3 & 0.7 & 0.05 & 92 & 5 \\
  4 & 0.7 & 0.05 & 360 & 5 \\
  5 & 0.3 & 0.01 & 360 & 10 \\
  \hline
  \end{tabular}
  \caption{Parameters used for training a $15\times 15$ SOM on 100,000 objects. \label{tab:training}} 
  \end{table}

\begin{figure}
  \includegraphics{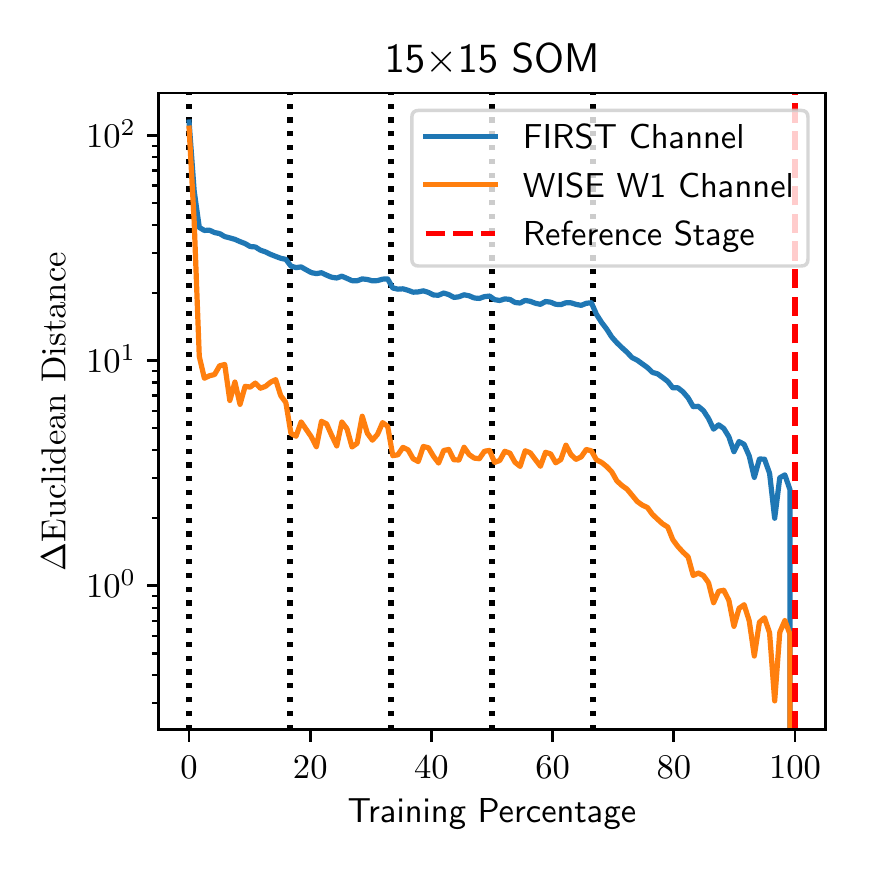}
  \caption{The change of the Euclidean distance of the SOM throughout training relative to its final state. We highlight the two SOM \revision{channels} independently. The vertical black dotted lines represent a \revision{change in the training stage as described in Table~\ref{tab:training}}, and the vertical red dashed line is the selected reference point. \label{fig:learning}}
\end{figure}

In Figure~\ref{fig:learning} we show the relative change of a SOM during its training phase across these five learning stages.
Intermediary SOMs were output by PINK at 5\% intervals throughout training during each stage.
We calculate the Euclidean distance (Equation~\ref{eq:ed}) between the final map, and the maps produced at each intermediary stage there after.
During the initial stage, where the broad layout of the SOM is established, there is a high rate of change in the Euclidean distance statistic.
There is also a high rate of change over time starting at roughly the 65\% point.
This is the point where we begin refining the map now that the general shape is well established.
Other learning stages exhibit relatively smaller changes.
The oscillating behaviour towards the end of the training is an indicator that individual steps during a single iteration end in more or less the initial state and therefore the map has converged.
\revision{To assess this behaviour we calculate the total intensity of each image cube and found the median value to be $\sim150$. The relative change of the entire SOM surface during the last training stages was $\sim10$. Hence, the change per neuron is approximately $(10/225)/150 = 0.03\%$ the median total image cube intensity. }
Although additional training stages could be made by decreasing the learning rate and \revisiontwo{region of influence of the neighborhood function} further, any improvements would be small and unnecessary for our purposes.

Experiments were run across multiple compute nodes, each equipped with four NVIDIA Tesla P40 GPUs, 60 CPU cores and a total of 128\,GB main memory.

\subsection{Sampling the Density of RGZ Labels}

An initial experiment performed using the PINK produced SOM was to assess how well the individual neurons do at separating \revisiontwo{object} labels from the labeled RGZ data-set. In a qualitative sense, a visual inspection of the outputted \revisiontwo{SOM lattice} across all channels (as we show later in Section~\ref{sec:results}) does show an evolution in morphologies. We produce a measure of the label distribution across individual neurons using the set of high consensus level RGZ \revisiontwo{objects}. For a properly trained SOM, there should be a representative prototype for each \revisiontwo{training object} image upon its \revisiontwo{lattice}. Hence, distributing \revisiontwo{object} labels to their corresponding prototype should result in a clustering of labels, particularly if the labels themselves are robust.  

The likelihood matrix (Equation~\ref{eq:ed-to-l-1}) of each of the 7,464 high consensus labeled \revisiontwo{objects} was used to distribute labels to neurons. \revision{These \revisiontwo{objects} are a subset of those used to train the SOM with.} To account for uncertainty in an \revisiontwo{object's} position, 100 realizations (with repetition allowed) of a. \revisiontwo{object's} position were drawn following its probabilities \revision{contained in its likelihood matrix}, where each realized position would receive a copy of the \revisiontwo{object's} label. Further, as the relative change of morphologies across adjacent neurons should be gradual, in principle distributing RGZ classifications in this manner should strengthen the clustering of labels and build a more reliable discrete probability function of each prototype's complexity. Prototype neurons with strong clustering of a labeled classification can be considered as reliable examples of that class. More complex prototype neurons should exhibit a spread of labels, particularly for the feature that is ambiguous or where the set of associated \revisiontwo{object} CLs are below one.     

\subsection{Label Transfer Experiment}

We conducted some experiments to check how well the user derived labels can be transferred using the dimensionality reduction method provided by PINK.
To analyse the prediction quality, the classification task based on the labels described in Table~\ref{tab:label_counts} has been split into independent regression tasks. 
Instead of predicting the combined number of components and number of peaks, those values are estimated separately.
Therefore the knowledge transfer experiments have been executed on both sets of labels. We explicitly treat this as a regression problem in order to characterise the ambiguity of an \revisiontwo{object} within the label itself. Consider an \revisiontwo{object} that is in an intermediary stage between one and two components. Its classification of either discrete label is dependent on the sigma contour level chosen \revisiontwo{to separate or highlight features within an object's image} and, in the case of RGZ, the subjective preference of the citizen scientist. A regressed value may be more appropriate to highlight the fuzzy or uncertain `classification' of an \revisiontwo{object}. 

The trained SOM is first used to calculate a heat-map for all objects with labels as described in Subsection~\ref{subsection:heatmaps}.
Those heat-maps are considered as input features for the regression task, as they encode the morphological characteristics of the objects through their distances to the learned prototypes.
By performing $k$-fold cross validation, a certain number of objects with labels is omitted when building up the regression model and the predictions are calculated for the omitted objects.
This allows for a proper prediction on all $7,464$ objects without having them being part of the used training data-set.
As we used a quantile regression forest, we can not only evaluate the mean over all ensemble members but can check the proper distribution of predictions, which in addition allows us to compute something equivalent to a consensus level.

\section{Results}
\label{sec:results}

\subsection{Visualisation of Neurons and Similarity}

Once the PINK software is applied to the training data-set it outputs the trained set of weights. Visualizing these weights allowed us to inspect the features in the images that the SOM has learnt to be distinguishing. When trained against image cubes with multiple planes or channels, there will be a corresponding channel in the trained SOM surface.

\begin{figure}
  \centering
\includegraphics[width=\linewidth, trim=0cm 2cm 0cm 2cm, clip=true]{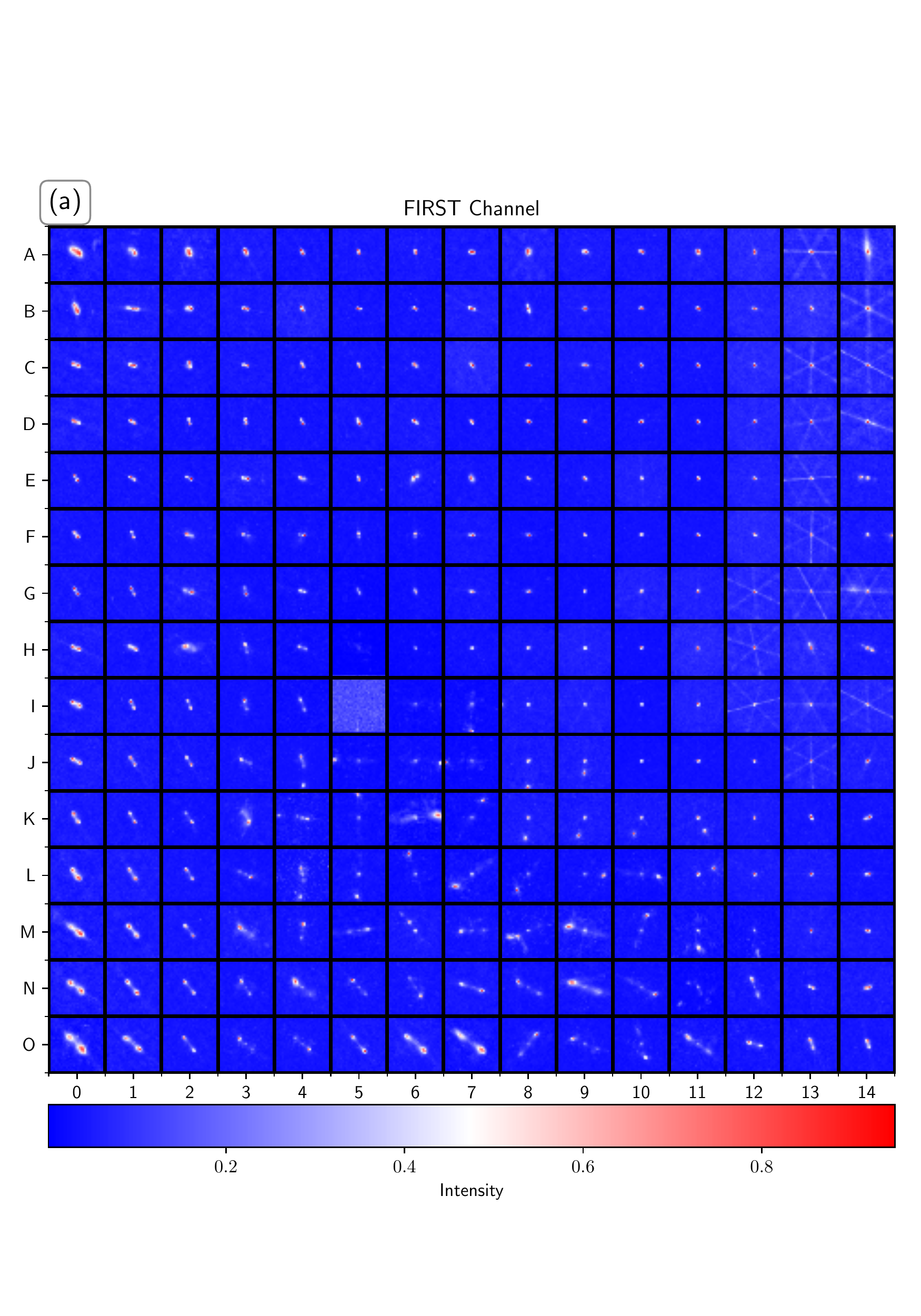}
\includegraphics[width=\linewidth, trim=0cm 2cm 0cm 0cm, clip=true]{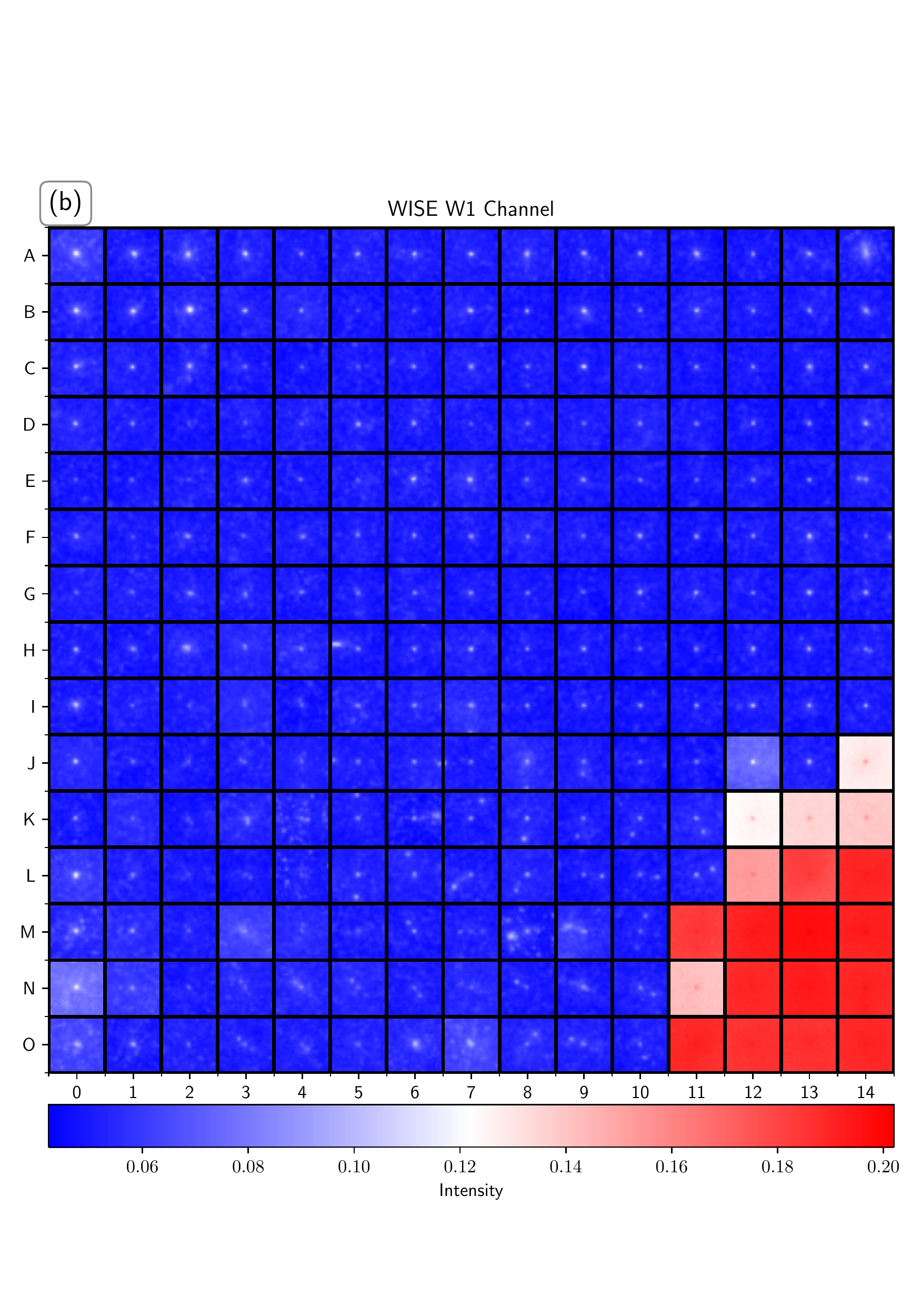}
\caption{\revision{We show the FIRST and WISE W1 channels of the SOM as the top \textit{(a)} and bottom \textit{(b)} panels respectively. }
The solid black horizontal and vertical lines separate  adjacent neurons on the SOM surface. \label{fig:FIRST_Norm} \label{fig:WISE_Norm}}
\end{figure}

We show in Figure~\ref{fig:FIRST_Norm} a visualization of the SOM trained by PINK using the pre-processed FIRST and WISE image cubes data described in Section~\ref{sec:preprocessing}.
This particular experiment trained a grid of $15\times15$ of neurons.
Within these 225 neurons, there is a clear change of source morphologies.

\revision{Within Figure~\ref{fig:FIRST_Norm}a, the neurons in the region bound by coordinates (A, 11), (A, 14), (I, 14) and (I, 9) contain unresolved point sources, including those with visible artefacts caused by the uncleaned VLA point spread function}. The remaining neurons consist of extended \revisiontwo{objects}. Objects with a clear, well defined separation between two components are contained to the lower left hand corner \revision{focused around position (M, 2) with an approximate bounding radii of 2 neurons}. The upper half of the map is made up of objects which exhibit islands of pixels connecting different sets of peaked pixels. Neurons in the upper left \revision{quadrant of the SOM} are sources which appear to exhibit brighter peak pixels, more diffuse structure and indication of a separate component than those in the upper right corner. The \revision{region bound by (I, 7), (I, 8), (K, 11) and (K, 5)} of the map shows neurons with two point source like components separate from one another in both the FIRST and WISE channels. The lower right corner of the WISE channel contain mostly image cubes which appear to have a calibration issue in their data. 

\begin{figure}
\includegraphics[width=\linewidth, clip=true, trim=1.5cm 0cm 0.5cm 1.cm]{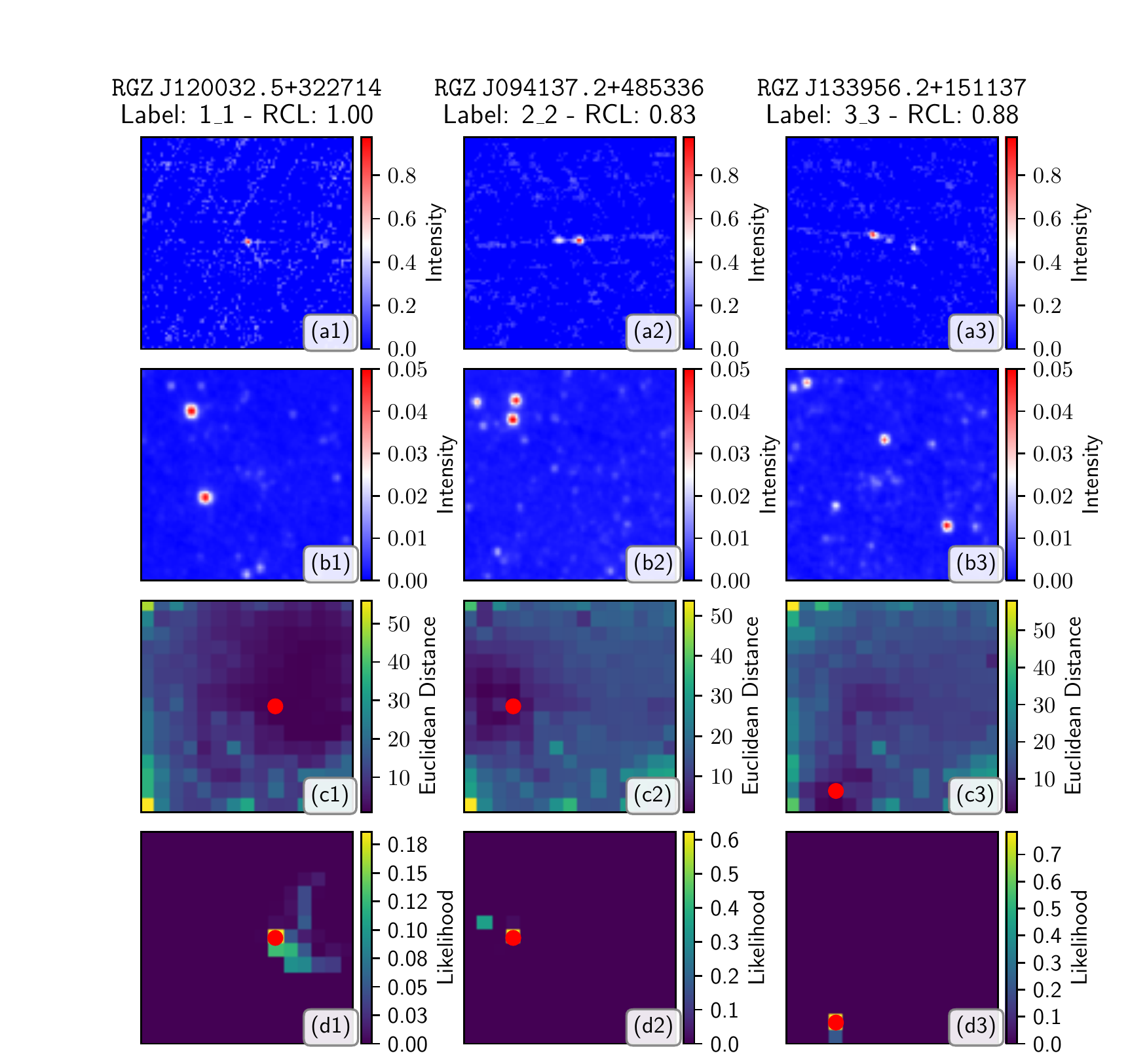}
\caption{The FIRST \textit{(row a)}  and WISE \textit{(row b)}  channels of the input cube supplied to PINK of objects \texttt{RGZ\,J120032.5+322714}, \texttt{RGZ\,J094137.2+485336} and \texttt{RGZ\,J133956.2+151137} with pre-processing stages applied. Under the object name of each column we include the RGZ designated label and the radio consensus level, abbreviated as RCL. \textit{(row c)} The image similarity matrix produced by PINK for rows (a) and (b) using the SOM presented in Figure~\ref{fig:FIRST_Norm}-\ref{fig:WISE_Norm}. \textit{(row d)} The likelihood matrix produced using row (c) and a $\psi=10$. The red circle in rows (c) and (d) denote the corresponding neuron from Figure~\ref{fig:FIRST_Norm} PINK judged to be most similar. \label{fig:heatmap}}
\end{figure}
%
%

\begin{figure*}[h!]
  \centering
  \includegraphics[width=\linewidth, clip=true, trim=3cm 0cm 3cm 3cm]{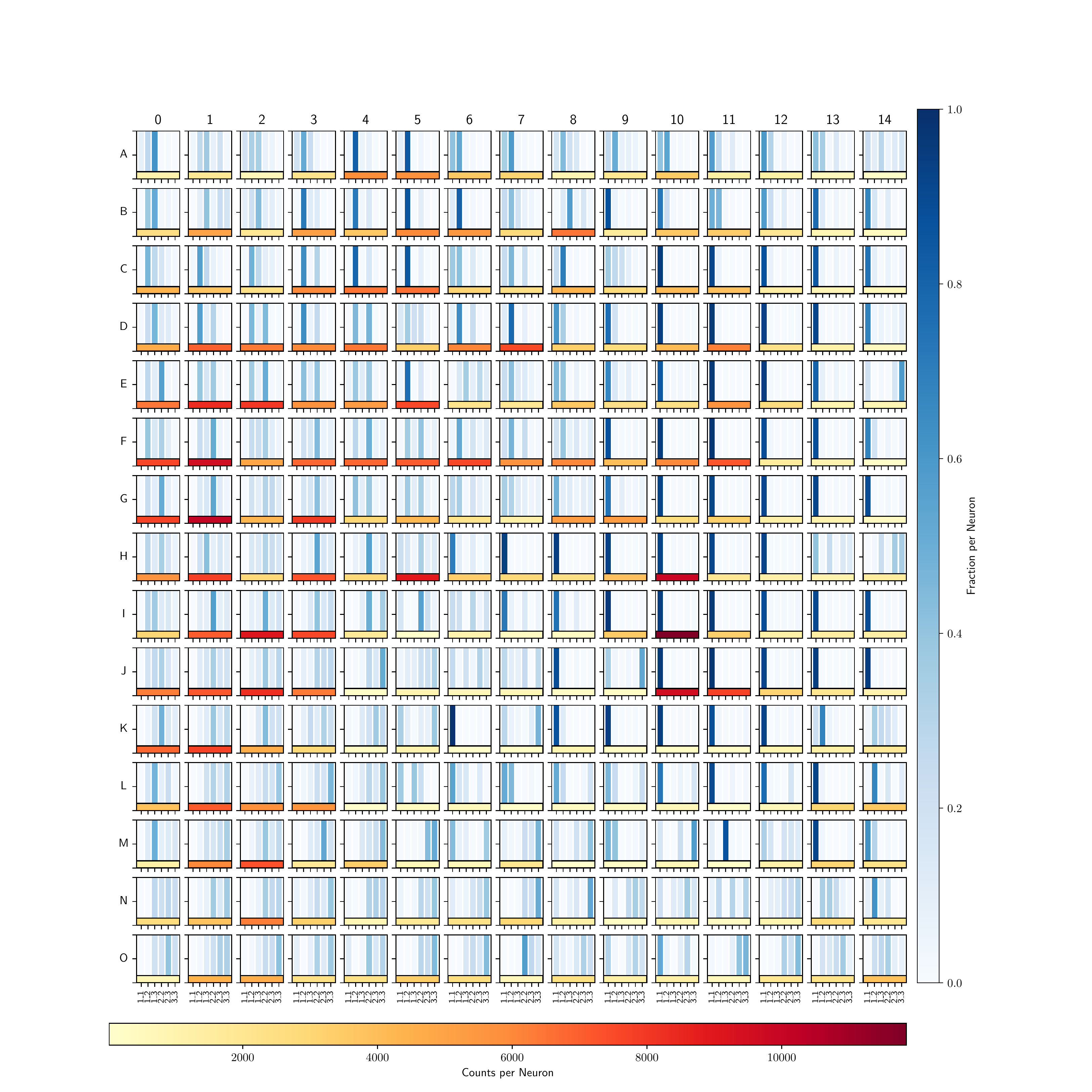}
  \caption{The distribution of labels of the training \revisiontwo{objects} outlined in Table~\ref{tab:label_counts} across the SOM \revisiontwo{lattice} presented in Figure~\ref{fig:FIRST_Norm}. Each histogram represents a single trained neuron, and the vertical colour bars of each label represents its contribution to the total set of labels of that neuron. \revision{Labels were distributed in a Monte-Carlo fashion across 100 realisation where their positions were randomly selected using each \revisiontwo{objects} likelihood matrix. } The horizontal bar in the lower half of each histogram encodes the number of \revisiontwo{objects} placed in the corresponding neuron across all realisations. A redder colour represents a higher number of labels in that neuron. The vertical bars in each histogram correspond to `1\_1',  `1\_2', `1\_3', `2\_2', `2\_3' and `3\_3' RGZ labels. \label{fig:label_dist_mc}}
\end{figure*}

 We present in Figure~\ref{fig:heatmap} examples of sources of different RGZ classifications with increasing complexities being mapped to the \revisiontwo{trained SOM lattice} shown in Figure~\ref{fig:FIRST_Norm}. We include the pre-processed FIRST and WISE images of \texttt{RGZ\,J120032.5+322714}, \texttt{RGZ\,J094137.2+485336} and \texttt{RGZ\,J133956.2+151137} which were provided to PINK. We also include in accompanying panels the corresponding Euclidean distance similarity matrix produced by PINK and the  likelihood matrix constructed using Equation~\ref{eq:ed-to-l-1} evaluated with a stretching parameter $\psi=10$. Both measures show distinct 
%
%
regions of activation in the reduced feature space. For the more complex labels, the likelihood matrices have a more compact region of activation, indicating a more conclusive prototype has been identified by PINK. For example, the simple object \texttt{RGZ\,J120032.5+322714}, whose RGZ label is `1\_1' and has a radio consensus level of 1.0, has a slightly larger region of activation. This region on the \revisiontwo{SOM lattice} smoothly transforms from simple point source structures to more complex morphologies. Since these are smooth changes, there is some ambiguity as to the most similar prototype, particularly as the majority of pixels in the input FIRST radio continuum image have been masked out as part of the pre-processing stages.

\subsection{RGZ Label Density}

We present in Figure~\ref{fig:label_dist_mc} the distribution of labels across the corresponding neuron grid for the SOM shown in Figure~\ref{fig:WISE_Norm}. 
\revision{Labels were distributed across neurons using the likelihood matrix of each source, created with $\psi=10$, and 100 realisations. } 
As a vertical color bar we highlight the discrete probability distribution function of labels for each segment on the SOM. 

Visualizing the density of labels in this manner, it is clear that there is evolution of morphologies among the neurons. The simpler classes with only a single component are clustered well in the upper half of the map, particularly around the (I,\,10) region. This is especially true for the `1\_1' label which show little activation to the left of column six.

Similarly, there are regions of ambiguity that can easily be distinguished. For instance, neurons surrounding (E,\,4) show activation of multiple labels with two peaks  (`1\_2' and `2\_2'). \revisiontwo{Object}s which PINK has deemed to be similar to these prototypes carry with them more difficulty to consistently classify the number of components. The neurons at positions (A,\,4) and (E,\,4) both appear to strongly preference one of (A,\,4)'s ambiguous labels. Similar behavior can also be seen for neurons between (N,\,1) to (N,\,8), where a variety of labels with two and three components show activation which gradually morphs into the activation of the `3\_3' label.  

It could therefore be considered that these regions where multiple labels are showing signs of activation are ambiguous prototypes and represent some type of intermediary stage in terms of morphology. The smooth transitions of features learnt by PINK have been able to capture these ambiguous \revisiontwo{object} prototypes.  

Included in Figure~\ref{fig:label_dist_mc} as a horizontal color bar is an indication of the number of labels associated with each neuron. This approach of visualizing the SOM gives an indication of the relative importance of each prototype morphology, particularly when considering the frequency of the labeled morphologies in the training data-set. Point source like objects (corresponding to a `1\_1' label) are concentrated around the (I,\,10) neuron, with counts above 12,000. Neurons whose discrete probability density function indicate more complex morphologies tend to have lower per neuron counts. Although this is somewhat explained by their lower fraction in the labeled data-set, an additional component is that PINK has required more neurons to accurately capture the gradual change in shapes among their collective set of morphologies. This thereby spreads the counts of those labels over a larger set of neuron locations. 

\subsection{Label Transfer}

\begin{figure}
\includegraphics[width=\linewidth, clip=true, trim=0cm 0cm 0cm 0cm]{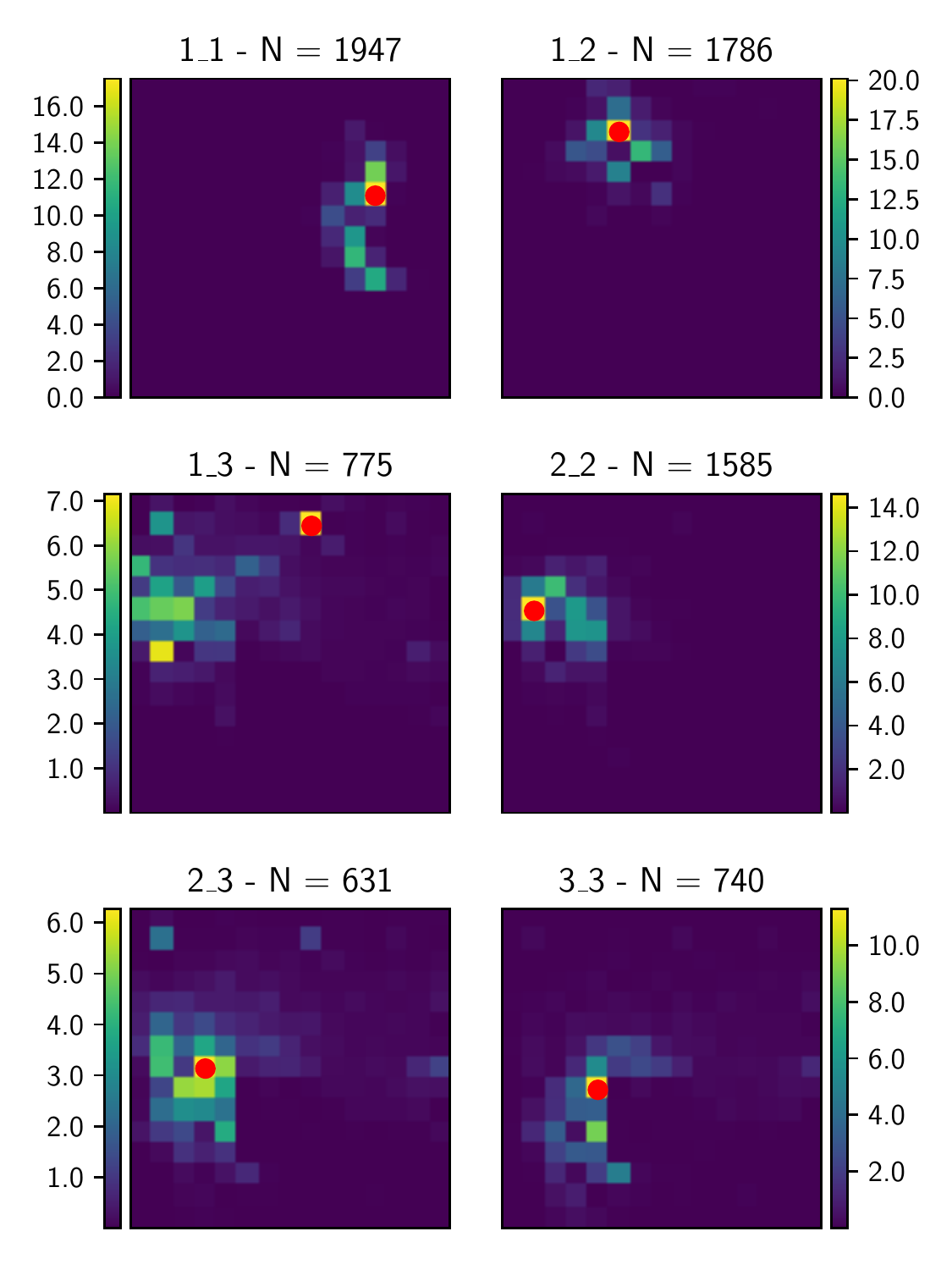}
\caption{The median likelihood matrix of each class of sources in the high consensus level training set, normalized for their total pixels values sum to one. We include the label and the total number of \revisiontwo{objects} with that label as a title for each panel. The overlaid red circle in each panel represents the neuron with the largest median probability across the surface. \label{fig:median_prob}}
\end{figure}

\revisiontwo{By grouping the likelihood matrices based on the corresponding RGZ label, an average or median likelihood matrix can be constructed for each class. These aggregate matrices highlight the typical set of neurons active across the SOM lattice for each label class. } We show in Figure~\ref{fig:median_prob} the median likelihood matrix (calculated using $\psi=10$) for each class of sources when projected onto the SOM \revisiontwo{lattice} shown in Figure~\ref{fig:FIRST_Norm}. Across the six classes, there is a clear separation between the regions of high likelihood and the gradient surrounding them. These structures can be used to make a prediction of an \revisiontwo{object's} corresponding label.


\begin{figure*}
  \centering
  \includegraphics[width=\linewidth]{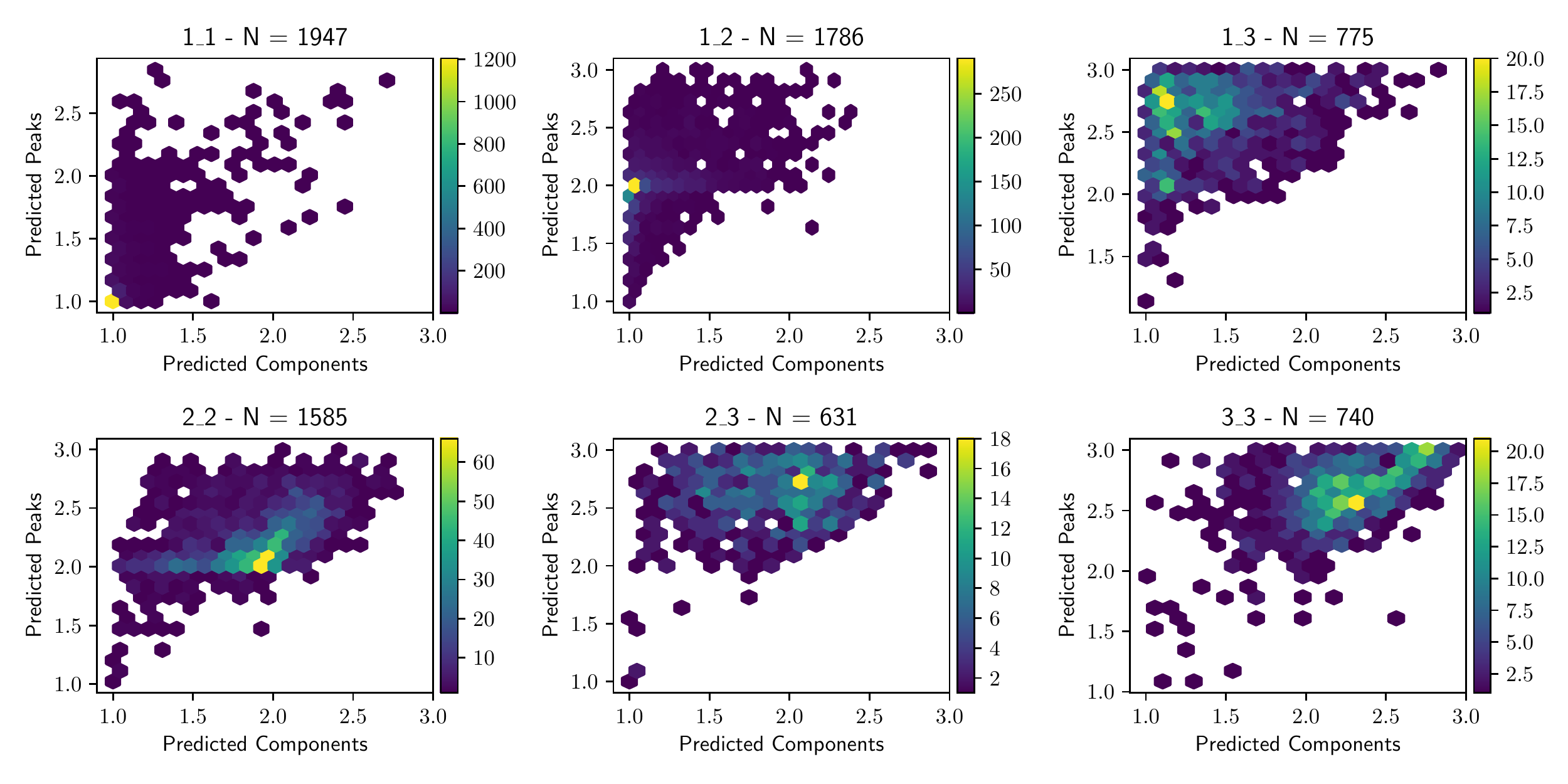}
  \caption{Two dimensional histograms highlighting the predicted number of components and classes made by the \texttt{RandomForestRegressor} for each \revisiontwo{object} grouped by their corresponding RGZ classifications. \revision{The intersection of each axis would produce an equivalent RGZ `$N_C\_N_P$' style label. } \label{fig:rfr_histograms}}
\end{figure*}

As outlined in Section~\ref{sec:rfr_outline}, we trained quantile random forest classifiers on the likelihood matrices of each labeled RGZ \revisiontwo{object} to predict their classified features. We independently constructed two \texttt{RandomForestRegressor}s to predict the number of components and peaks for each labelled \revisiontwo{object}, and show in Figure~\ref{fig:rfr_histograms} the results of their predictive power grouped by the RGZ designated label. The intersection of \revisiontwo{the predicted number of components axis and the predicted number of peaks axis can be considered the equivalent RGZ `$N_C\_N_P$' style label}. However, as described in Section~\ref{sec:rfr_outline}, we are taking the mean of the predicted labels across each of the independently trained decision trees, allowing for a `fuzzy' label to be considered. Constructing labels in this manner allows for a mechanism to encode the ambiguity of an object's classification. 

The predicted regressed values shown in Figure~\ref{fig:rfr_histograms} for \revisiontwo{objects} whose RGZ classification specifies a single component each show that the bin with the most counts agree with the corresponding label. For the `1\_1' and `1\_2' there is a clear separation between the bin with the maximum number of counts and their surrounding cells. Regressed values for objects with a `1\_3' RGZ classification are more spread across the surface. The location of the maximum bin (which is located approximately at 2.75) also suggests that there is a fraction of voting trees who favor a lower number of peaks.

Following this, there is also a spread of predicted components and peaks made by the trained \texttt{RandomForest\-Regressor} for more complex RGZ labels. For \revisiontwo{objects} classified as having two components, there is clear structure among the two-dimensional histograms which broadly agree with the corresponding RGZ label. The spread across the \revisiontwo{lattice} can be seen as natural, as the increased degree of complexity in an \revisiontwo{object's} morphology would lead to a larger disagreement in the cohort of decision trees. This behaviour is also exhibited as a lower consensus level for these RGZ \revisiontwo{objects} (Figure~\ref{fig:consensus_level}).     

The transition and ambiguity between `2\_3' and `3\_3' can also be seen in their corresponding histograms in Figure~\ref{fig:rfr_histograms}. A small but clear set of bins in the `2\_3' histogram is beginning to form in the approximate location of `2\_2.5', which is where the peak bin is located for `3\_3'. Between these histograms, however, the region of activation is trending to higher complexity for the RGZ `3\_3' label. 

\section{Discussion}
\label{sec:discussion}

\subsection{Distance from Labeled Features}

\begin{figure*}
\includegraphics[width=\linewidth]{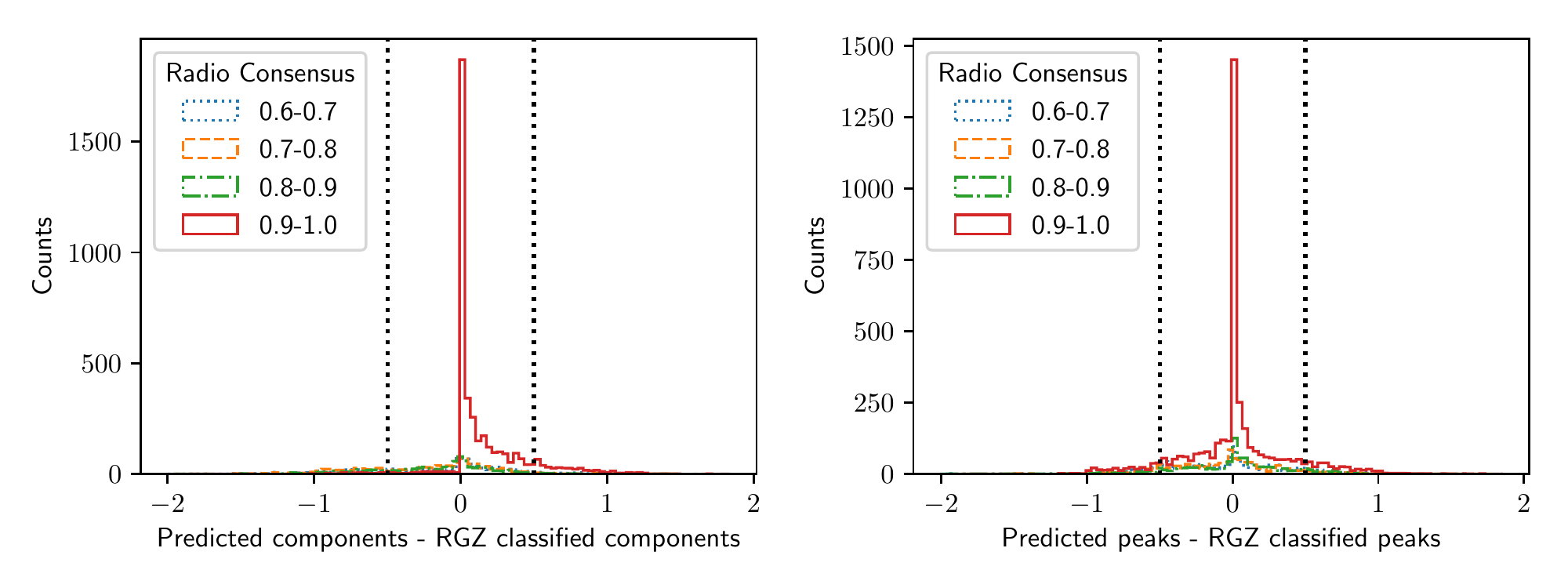}
\caption{The difference between the number of components and peaks predicted by a \texttt{RandomForestRegressor} trained using the likelihood matrix and the number of components or peaks classified by the RGZ citizen scientists. The results have been grouped by the radio CL. The dotted vertical lines represent the regions that would be rounded to zero. \label{fig:mse_hist}}

  \centering\tiny
  \begin{tabular}{ll|rrrrrrr}
    \toprule
     \multicolumn{9}{c}{Predicted Component Performance} \\
     \toprule
     \textbf{RGZ} & \textbf{Prediction}&  1\_1 &  1\_2 &  1\_3 &  2\_2 &  2\_3 &  3\_3 &  Summary \\
\textbf{Radio CL} & \textbf{Result} &      &      &      &      &      &      &          \\
\midrule
\midrule
\textbf{0.6-0.7} & \textbf{Incorrect} &    23 / 0.90 / 0.22 &    17 / 0.77 / 0.17 &   28 / 0.85 / 0.19 &    61 / 0.69 / 0.14 &   32 / 0.70 / 0.16 &  214 / 0.90 / 0.31 &   375 / 0.84 / 0.27 \\
        & \textbf{Correct} &   111 / 0.10 / 0.13 &    84 / 0.14 / 0.16 &   22 / 0.19 / 0.12 &   339 / 0.19 / 0.13 &  146 / 0.22 / 0.14 &   65 / 0.36 / 0.10 &   767 / 0.19 / 0.15 \\
\textbf{0.7-0.8} & \textbf{Incorrect} &    10 / 1.15 / 0.35 &    10 / 0.95 / 0.23 &   21 / 0.80 / 0.26 &    91 / 0.67 / 0.13 &   51 / 0.67 / 0.13 &  225 / 0.89 / 0.27 &   408 / 0.82 / 0.26 \\
        & \textbf{Correct} &    56 / 0.06 / 0.09 &    51 / 0.18 / 0.15 &   21 / 0.20 / 0.15 &   403 / 0.17 / 0.12 &  170 / 0.19 / 0.14 &   67 / 0.31 / 0.11 &   768 / 0.18 / 0.14 \\
\textbf{0.8-0.9} & \textbf{Incorrect} &     3 / 1.11 / 0.06 &    12 / 0.74 / 0.24 &   17 / 0.86 / 0.31 &    90 / 0.67 / 0.12 &   40 / 0.65 / 0.13 &  111 / 0.96 / 0.29 &   273 / 0.81 / 0.27 \\
        & \textbf{Correct} &    58 / 0.05 / 0.10 &    66 / 0.14 / 0.14 &   25 / 0.27 / 0.15 &   422 / 0.17 / 0.13 &  135 / 0.21 / 0.13 &   36 / 0.32 / 0.11 &   742 / 0.18 / 0.14 \\
\textbf{0.9-1.0} & \textbf{Incorrect} &    27 / 0.74 / 0.26 &   201 / 0.73 / 0.18 &  223 / 0.81 / 0.27 &    30 / 0.69 / 0.12 &   19 / 0.70 / 0.14 &   21 / 0.96 / 0.31 &   521 / 0.77 / 0.24 \\
        & \textbf{Correct} &  1659 / 0.03 / 0.07 &  1345 / 0.12 / 0.13 &  418 / 0.21 / 0.14 &   149 / 0.19 / 0.13 &   38 / 0.19 / 0.12 &    1 / 0.26 / 0.00 &  3610 / 0.09 / 0.13 \\
\bottomrule
\textbf{Summary} & {} &  1947 / 0.06 / 0.18 &  1786 / 0.20 / 0.26 &  775 / 0.44 / 0.35 &  1585 / 0.26 / 0.23 &  631 / 0.31 / 0.24 &  740 / 0.78 / 0.36 &  7464 / 0.27 / 0.33 \\
\bottomrule
\toprule
\multicolumn{9}{c}{Predicted Peak Performance} \\
\toprule
\textbf{RGZ} &   \textbf{Prediction}&  1\_1 &  1\_2 &  1\_3 &  2\_2 &  2\_3 &  3\_3 &  Summary \\
\textbf{Radio CL} & \textbf{Result} &      &      &      &      &      &      &          \\
\midrule
\textbf{0.6-0.7} & \textbf{Incorrect} &    40 / 0.98 / 0.30 &    16 / 0.67 / 0.12 &   21 / 0.72 / 0.16 &    86 / 0.65 / 0.11 &   55 / 0.72 / 0.22 &   80 / 0.75 / 0.29 &   298 / 0.74 / 0.24 \\
        & \textbf{Correct} &    94 / 0.12 / 0.14 &    85 / 0.15 / 0.14 &   29 / 0.28 / 0.13 &   314 / 0.21 / 0.15 &  123 / 0.27 / 0.15 &  199 / 0.26 / 0.14 &   844 / 0.22 / 0.15 \\
\textbf{0.7-0.8} & \textbf{Incorrect} &    16 / 1.06 / 0.39 &    11 / 0.75 / 0.16 &   16 / 0.80 / 0.28 &    87 / 0.68 / 0.14 &   57 / 0.71 / 0.21 &   78 / 0.69 / 0.22 &   265 / 0.72 / 0.23 \\
        & \textbf{Correct} &    50 / 0.08 / 0.13 &    50 / 0.16 / 0.15 &   26 / 0.25 / 0.11 &   407 / 0.17 / 0.14 &  164 / 0.23 / 0.14 &  214 / 0.23 / 0.15 &   911 / 0.19 / 0.15 \\
\textbf{0.8-0.9} & \textbf{Incorrect} &    10 / 1.00 / 0.29 &    12 / 0.65 / 0.14 &   12 / 0.79 / 0.20 &    90 / 0.64 / 0.10 &   50 / 0.75 / 0.30 &   35 / 0.77 / 0.28 &   209 / 0.71 / 0.23 \\
        & \textbf{Correct} &    51 / 0.08 / 0.11 &    66 / 0.12 / 0.13 &   30 / 0.23 / 0.14 &   422 / 0.17 / 0.14 &  125 / 0.26 / 0.14 &  112 / 0.23 / 0.15 &   806 / 0.18 / 0.15 \\
\textbf{0.9-1.0} & \textbf{Incorrect} &   215 / 0.83 / 0.27 &   203 / 0.68 / 0.13 &  218 / 0.77 / 0.22 &    25 / 0.67 / 0.11 &   22 / 0.68 / 0.14 &    5 / 0.85 / 0.29 &   688 / 0.76 / 0.22 \\
        & \textbf{Correct} &  1471 / 0.05 / 0.11 &  1343 / 0.14 / 0.14 &  423 / 0.25 / 0.14 &   154 / 0.15 / 0.14 &   35 / 0.26 / 0.15 &   17 / 0.25 / 0.18 &  3443 / 0.12 / 0.14 \\
\bottomrule
\textbf{Summary} & {} &  1947 / 0.17 / 0.32 &  1786 / 0.21 / 0.23 &  775 / 0.43 / 0.30 &  1585 / 0.27 / 0.23 &  631 / 0.39 / 0.28 &  740 / 0.37 / 0.29 &  7464 / 0.27 / 0.29 \\
\bottomrule
    \end{tabular}
    
\captionof{table}{An overview of the predictive performance of the two \texttt{RandomForestRegressor}s predicting the number of components (top) and number of peaks (bottom) which have been trained using the likelihood matrix produced by PINK. The results have been broken down by the RGZ Radio CL, the RGZ label and whether the prediction result is correct if a hard classification scheme is used. For each intersection (including the summary column and row) we include the number of \revisiontwo{objects} in that subset, their average absolute distance from zero (a perfect prediction) and the standard deviation of the absolute distance from zero. \label{tab:rgz_rfr_predictor}}
\end{figure*}

The use of fuzzy labels (eg. continuous predicted features opposed to discrete labels) allows for an assessment of the distance between the predicted number of components and peaks from the classified number to be made, a comparison which we present in Figure~\ref{fig:mse_hist}. For \revisiontwo{objects} where the entire cohort of decision trees unanimously voted for a single feature which matched the corresponding RGZ label, the distance would be zero. As \revisiontwo{objects} start to exhibit more complex features, ambiguity among the decision trees would trend them away from the zero point. 
\revision{Results of the predictive performance in subsequent discussion of the \texttt{RandomForest\-Regressor} are based on the results when each \revisiontwo{object} was in the testing segment of the $k$-fold cross validation method. }

We find through this comparison that the method of classifying to reduced feature space produced by the  similarity measure of PINK is a powerful predictor of an \revisiontwo{object's} label. \revision{For this test we produced a likelihood matrix for each source with $\psi=1$}. Both panels of Figure~\ref{fig:mse_hist} exhibit a strong feature centered on zero far above the residual error across all radio consensus levels. The slight positive bias seen in the comparison of predicted components to RGZ classified components originates from the large number of RGZ labels included in the training labeled data-set and the feature predictor never under-predicting the number of components. Although there appears to be a positive skew in the number of predicted components, this is explained simply by the \texttt{RandomForestRegressor} never predicting less than 1 component for \revisiontwo{objects} with `1\_1', `1\_2' or `1\_3' RGZ labels, which are collectively a large fraction of the labeled training data and have high CLs. There are no systematic biases when similar figures are made on a RGZ label basis.  

If the soft label scheme we adopt here is converted to a hard label boundary, similar to the RGZ labels themselves, the vertical dotted lines represents the \revisiontwo{region which would} be `rounded' to zero \revisiontwo{- a correct classification}. For the number of predicted components and peaks this encompasses \revision{79.0\% and 80.7\%} of the data-set, respectively. For reference, using the fuzzy labels, we compute the sum of the absolute distance from zero for this inner region to be \revision{751.4 and 920.3} for the components and peaks regressors. We note that this is not necessarily an indication that the training method is poor, as the training labels themselves carry uncertainty. 

\revisiontwo{A complete overview of the performance of the predictive component and peak \texttt{RandomForest\-Regressor}s is presented in Table~\ref{tab:rgz_rfr_predictor}, including the predictive performance across RGZ labels and CL subsets. }

\begin{figure*}
  \includegraphics[width=\linewidth]{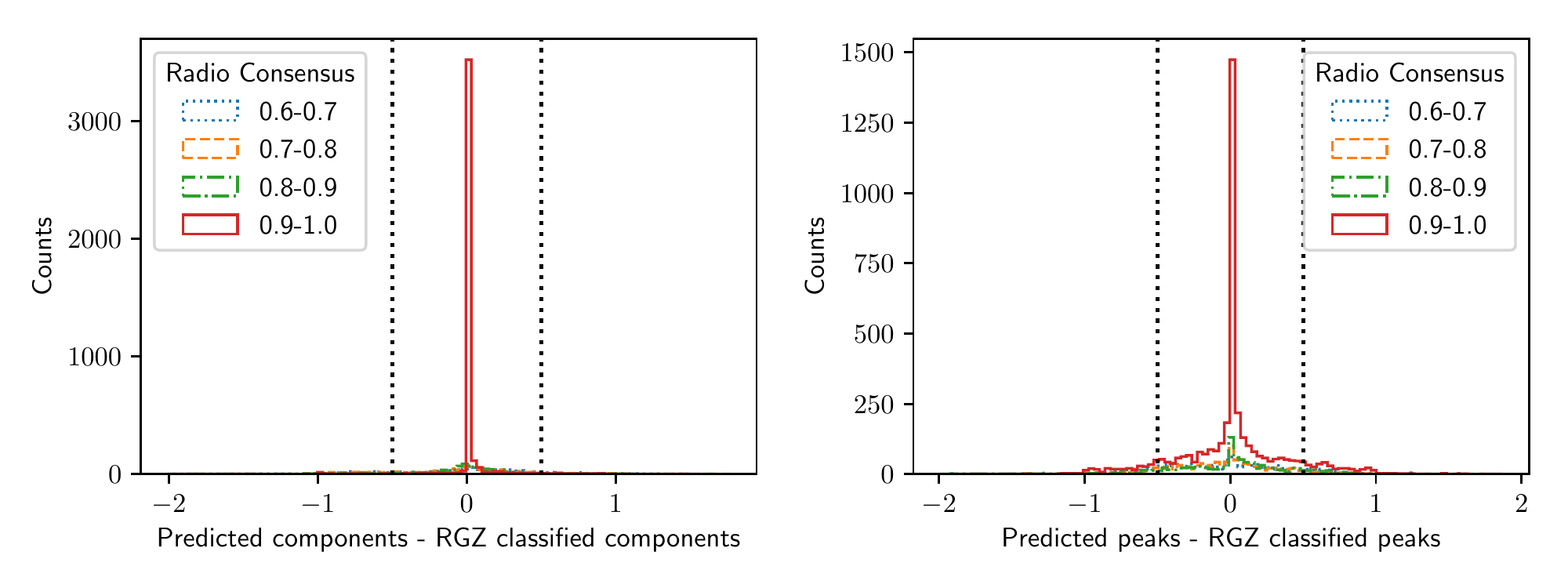}
  \caption{The difference between the number of components and peaks predicted by a \texttt{RandomForestRegressor} trained using the likelihood matrix supplemented with the radio CL and the number of components or peaks classified by the RGZ citizen scientists. The results have been grouped by the radio CL. The dotted vertical lines represent the regions that would be rounded to zero. \label{fig:mse_hist_cl} }

  \centering\tiny
  \begin{tabular}{ll|rrrrrrr}
    \toprule
     \multicolumn{9}{c}{Predicted Component Performance} \\
     \toprule
     \textbf{RGZ} & \textbf{Prediction} &  1\_1 &  1\_2 &  1\_3 &  2\_2 &  2\_3 &  3\_3 &  Summary \\
\textbf{Radio CL} & \textbf{Result} &      &      &      &      &      &      &          \\
\midrule
\textbf{0.6-0.7} & \textbf{Incorrect} &    31 / 0.91 / 0.26 &    48 / 0.86 / 0.22 &   46 / 0.97 / 0.26 &    20 / 0.64 / 0.12 &   20 / 0.62 / 0.12 &  198 / 0.80 / 0.25 &   363 / 0.82 / 0.26 \\
        & \textbf{Correct} &   103 / 0.15 / 0.14 &    53 / 0.21 / 0.16 &    4 / 0.38 / 0.12 &   380 / 0.17 / 0.13 &  158 / 0.23 / 0.14 &   81 / 0.34 / 0.11 &   779 / 0.20 / 0.14 \\
\textbf{0.7-0.8} & \textbf{Incorrect} &    14 / 1.05 / 0.39 &    36 / 0.86 / 0.23 &   35 / 1.00 / 0.28 &    22 / 0.64 / 0.13 &   24 / 0.62 / 0.10 &  199 / 0.78 / 0.19 &   330 / 0.80 / 0.23 \\
        & \textbf{Correct} &    52 / 0.12 / 0.13 &    25 / 0.19 / 0.14 &    7 / 0.27 / 0.14 &   472 / 0.15 / 0.12 &  197 / 0.20 / 0.14 &   93 / 0.33 / 0.12 &   846 / 0.18 / 0.14 \\
\textbf{0.8-0.9} & \textbf{Incorrect} &     9 / 0.84 / 0.23 &    42 / 0.77 / 0.19 &   36 / 0.98 / 0.26 &    11 / 0.57 / 0.06 &   20 / 0.67 / 0.12 &  106 / 0.83 / 0.22 &   224 / 0.82 / 0.23 \\
        & \textbf{Correct} &    52 / 0.13 / 0.12 &    36 / 0.24 / 0.16 &    6 / 0.36 / 0.11 &   501 / 0.12 / 0.11 &  155 / 0.17 / 0.12 &   41 / 0.32 / 0.10 &   791 / 0.15 / 0.13 \\
\textbf{0.9-1.0} & \textbf{Incorrect} &     2 / 0.58 / 0.05 &    33 / 0.75 / 0.14 &   50 / 0.79 / 0.22 &    38 / 0.82 / 0.18 &   15 / 0.76 / 0.19 &   21 / 1.16 / 0.48 &   159 / 0.83 / 0.28 \\
        & \textbf{Correct} &  1684 / 0.00 / 0.02 &  1513 / 0.02 / 0.06 &  591 / 0.02 / 0.07 &   141 / 0.16 / 0.13 &   42 / 0.16 / 0.12 &    1 / 0.29 / 0.00 &  3972 / 0.02 / 0.06 \\
\bottomrule
\textbf{Summary} & {} &  1947 / 0.04 / 0.17 &  1786 / 0.10 / 0.25 &  775 / 0.22 / 0.40 &  1585 / 0.18 / 0.18 &  631 / 0.25 / 0.20 &  740 / 0.67 / 0.31 &  7464 / 0.18 / 0.30 \\
\bottomrule
\toprule
\multicolumn{9}{c}{Predicted Peak Performance} \\
\toprule
\textbf{RGZ} &  \textbf{Prediction} &  1\_1 &  1\_2 &  1\_3 &  2\_2 &  2\_3 &  3\_3 &  Summary \\
\textbf{Radio CL} & \textbf{Result} &      &      &      &      &      &      &          \\
\midrule
\textbf{0.6-0.7} & \textbf{Incorrect} &    43 / 0.96 / 0.29 &    16 / 0.65 / 0.12 &   22 / 0.71 / 0.15 &    93 / 0.64 / 0.11 &   54 / 0.71 / 0.22 &   83 / 0.73 / 0.28 &   311 / 0.73 / 0.24 \\
        & \textbf{Correct} &    91 / 0.12 / 0.14 &    85 / 0.16 / 0.14 &   28 / 0.27 / 0.14 &   307 / 0.21 / 0.14 &  124 / 0.27 / 0.14 &  196 / 0.26 / 0.13 &   831 / 0.22 / 0.15 \\
\textbf{0.7-0.8} & \textbf{Incorrect} &    15 / 1.11 / 0.36 &    11 / 0.75 / 0.17 &   15 / 0.81 / 0.27 &    84 / 0.68 / 0.13 &   51 / 0.72 / 0.21 &   71 / 0.71 / 0.23 &   247 / 0.74 / 0.23 \\
        & \textbf{Correct} &    51 / 0.10 / 0.14 &    50 / 0.17 / 0.15 &   27 / 0.25 / 0.12 &   410 / 0.18 / 0.14 &  170 / 0.23 / 0.14 &  221 / 0.24 / 0.15 &   929 / 0.20 / 0.15 \\
\textbf{0.8-0.9} & \textbf{Incorrect} &    10 / 0.99 / 0.32 &    10 / 0.67 / 0.12 &   13 / 0.75 / 0.20 &    88 / 0.64 / 0.11 &   55 / 0.72 / 0.28 &   37 / 0.76 / 0.28 &   213 / 0.71 / 0.23 \\
        & \textbf{Correct} &    51 / 0.09 / 0.11 &    68 / 0.13 / 0.14 &   29 / 0.24 / 0.14 &   424 / 0.17 / 0.14 &  120 / 0.25 / 0.13 &  110 / 0.22 / 0.15 &   802 / 0.18 / 0.15 \\
\textbf{0.9-1.0} & \textbf{Incorrect} &   209 / 0.83 / 0.27 &   200 / 0.68 / 0.13 &  231 / 0.76 / 0.23 &    26 / 0.65 / 0.12 &   24 / 0.67 / 0.14 &    6 / 0.79 / 0.26 &   696 / 0.75 / 0.22 \\
        & \textbf{Correct} &  1477 / 0.05 / 0.11 &  1346 / 0.13 / 0.14 &  410 / 0.25 / 0.13 &   153 / 0.15 / 0.13 &   33 / 0.25 / 0.14 &   16 / 0.22 / 0.17 &  3435 / 0.11 / 0.14 \\
\bottomrule
\textbf{Summary} & {} &  1947 / 0.17 / 0.32 &  1786 / 0.21 / 0.23 &  775 / 0.43 / 0.30 &  1585 / 0.27 / 0.23 &  631 / 0.38 / 0.27 &  740 / 0.37 / 0.28 &  7464 / 0.27 / 0.29 \\
\bottomrule
    \end{tabular}
    
\captionof{table}{An overview of the predictive performance of the two \texttt{RandomForestRegressor}s predicting the number of components (top) and number of peaks (bottom) which have been trained using the likelihood matrix produced by PINK supplemented by the RGZ radio CL of each object. The results have been broken down by the RGZ Radio CL, the RGZ label and whether the prediction result is correct if a hard classification scheme is used. For each intersection (including the summary column and row) we include the number of \revisiontwo{objects} in that subset, their average absolute distance from zero (a perfect prediction) and the standard deviation of the absolute distance from zero.  \label{tab:rgz_rfr_cl_predictor}}
\end{figure*}

A powerful mechanism of the use of the \texttt{RandomForest\-Regressor} is that it is not tied directly to the use of the PINK produced similarity matrices, whether they be in the form of a Euclidean distance or likelihood. They can be supplemented with additional catalogue space information, such as redshifts or optical magnitudes. Incorporating this information can help to add additional constraints when making a prediction. As an example, we trained a \texttt{RandomForest\-Regressor} using both the likelihood matrices and the RGZ radio CLs of each \revisiontwo{object} as a single feature vector of length 226. We show in Figure~\ref{fig:mse_hist_cl} the difference between predicted and RGZ classified number of components and peaks. There is a clear improvement in the performance of the regressor when compared to Figure~\ref{fig:mse_hist}, particularly for the predicted number of components. When converted to a hard classification scheme, \revision{85.7\% and 80.7\%} of the \revisiontwo{objects} would have a correctly predicted number of components and peaks if the RGZ labels are accepted as ground truth. The sum of the absolute distance for the predicted number of components and peaks is \revision{503.2 and 919.5} respectively. The inclusion of the radio consensus labels has improved the predictive power of the number of components, while the number of peaks has remained similar to training with purely the PINK similarity measures. 

\revisiontwo{We include in Table~\ref{tab:rgz_rfr_cl_predictor} a complete overview of the predictive performance of the \texttt{RandomForest\-Regressor}s trained using the likelihood matix supplemented with each object's CL. These results are also broken down across RGZ labels and CL subsets. }

The addition of supplementary features is not limited to a single feature. In practice a large collection of data can be added, including sparse or incomplete types of data. For example, spectroscopic classifications from the Sloan Digital Sky Survey \citep[SDSS; ][]{2000AJ....120.1579Y} can be added to \revisiontwo{objects} if they are available without having to exclude others which do not have SDSS identifications. 

\revision{For both sets of \texttt{RandomForestRegressor}s trained using outputs of PINK the predictive performance of the `1\_3' and `3\_3' across most consensus levels was less than 50\%. These classes also happened to be the two of the smaller groups in our labeled data-set. As an initial test we restricted the size of each class to $631$ \revisiontwo{objects} (the number of \revisiontwo{objects} in the smallest group) and performed the same tests. We found that this balanced data-set performed in a consistent manner to the previous \texttt{RandomForestRegressor}s without any meaningful difference in its accuracy.} 
\revision{There are potentially two compounding effects which are contributing to the relatively poor performance of these classes. \revisiontwo{objects} with these labels are beginning to exhibit complex features that are more dependent on the subjective classification nature of the participating citizen scientists. Defining the number of components based on an image sigma threshold mechanism, particularly for \revisiontwo{objects} whose brightness is close to image sensitivity limits, are particularly susceptible to inconsistent responses. This type of behaviour is captured by the radio CL. We also suspect that this issue is being aggravated by there not being enough space on the \revisiontwo{SOM lattice} for these distinct complex features to distinguish themselves. Examining Figure~\ref{fig:median_prob} there is overlap in the activation regions of these complex classes. Experimenting with larger SOM sizes maintained the relative proportions of these regions. A more likely approach to improve performance of these classes is to produce a type of hierarchical SOM where data is segmented based on an initial layer, and additional SOMs are trained on these subsets separately in a manner similar to a growing hierarchical SOM \citep{dittenbach2000growing}. }

\subsection{Improvement from PINK}

A point to consider is whether the PINK method itself is an effective resource that helps the \texttt{RandomForest\-Regressor} predict \revisiontwo{object} labels. Given that preprocessing stages were employed to better emphasise important features, there is potential for the \texttt{RandomForestRegressor} to perform equally well at distinguishing important shapes and classifying morphologies.   

To explore this, we provided the 7,464 preprocessed image cubes used by PINK to the 
\texttt{RandomForest\-Regressor} class. These image cubes were flattened to a $2\times167\times167=55,778$ feature vector. We utilized the same experimental setup used in the previous section to produce Figures~\ref{fig:rfr_histograms} and~\ref{fig:mse_hist}, and performed the same number of cross validation tests.  

\begin{figure*}
  \includegraphics[width=\linewidth]{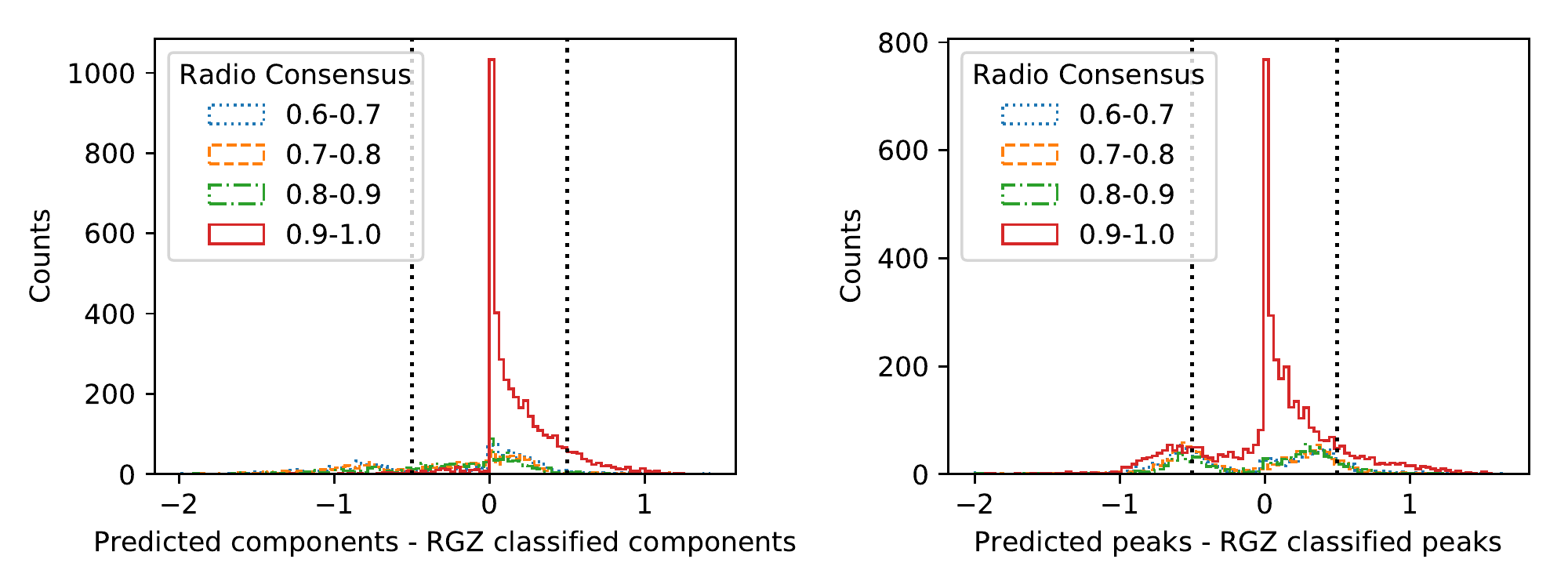}
  \caption{The difference between the predicted number of components and peaks against the number classified by the RGZ citizen scientists, where the results have been grouped by the radio CL. The \texttt{RandomForestRegressor}s has been trained using the pre-processed images provided to PINK. The dotted vertical lines represent the regions that would be rounded to zero. \label{fig:mse_hist_pink} }
  
  \tiny
  \begin{tabular}{ll|lllllll}
    \toprule
    \multicolumn{9}{c}{Predicted Component Performance} \\
    \toprule
      \textbf{RGZ}  &  \textbf{Prediction}  &          1\_1 &          1\_2 &         1\_3 &          2\_2 &         2\_3 &         3\_3 &      Summary \\
    \textbf{Radio CL} & \textbf{Result} &              &              &             &              &             &             &              \\
    \midrule
\textbf{0.6-0.7} & \textbf{Incorrect} &    28 / 0.84 / 0.26 &    16 / 0.73 / 0.19 &   13 / 0.83 / 0.22 &    46 / 0.67 / 0.13 &   35 / 0.71 / 0.14 &  279 / 1.00 / 0.29 &   417 / 0.91 / 0.29 \\
        & \textbf{Correct} &   106 / 0.12 / 0.14 &    85 / 0.16 / 0.14 &   37 / 0.23 / 0.14 &   354 / 0.19 / 0.12 &  143 / 0.20 / 0.12 &               None &   725 / 0.18 / 0.13 \\
\textbf{0.7-0.8} & \textbf{Incorrect} &    15 / 0.78 / 0.23 &    13 / 0.79 / 0.23 &   13 / 0.84 / 0.17 &    57 / 0.65 / 0.14 &   32 / 0.71 / 0.14 &  287 / 1.02 / 0.28 &   417 / 0.92 / 0.29 \\
        & \textbf{Correct} &    51 / 0.11 / 0.13 &    48 / 0.18 / 0.14 &   29 / 0.20 / 0.14 &   437 / 0.19 / 0.12 &  189 / 0.19 / 0.12 &    5 / 0.23 / 0.12 &   759 / 0.18 / 0.12 \\
\textbf{0.8-0.9} & \textbf{Incorrect} &     7 / 0.72 / 0.16 &    10 / 0.65 / 0.20 &   16 / 0.72 / 0.17 &    53 / 0.63 / 0.11 &   28 / 0.73 / 0.13 &  143 / 1.07 / 0.29 &   257 / 0.89 / 0.31 \\
        & \textbf{Correct} &    54 / 0.10 / 0.13 &    68 / 0.16 / 0.15 &   26 / 0.22 / 0.15 &   459 / 0.19 / 0.13 &  147 / 0.19 / 0.13 &    4 / 0.32 / 0.13 &   758 / 0.18 / 0.13 \\
\textbf{0.9-1.0} & \textbf{Incorrect} &    74 / 0.66 / 0.19 &   211 / 0.69 / 0.18 &  183 / 0.73 / 0.17 &    24 / 0.62 / 0.10 &    9 / 0.64 / 0.12 &   22 / 1.12 / 0.31 &   523 / 0.71 / 0.20 \\
        & \textbf{Correct} &  1612 / 0.10 / 0.12 &  1335 / 0.16 / 0.14 &  458 / 0.21 / 0.14 &   155 / 0.19 / 0.12 &   48 / 0.21 / 0.13 &               None &  3608 / 0.14 / 0.14 \\
\bottomrule
\textbf{Summary} & {} &  1947 / 0.14 / 0.20 &  1786 / 0.23 / 0.24 &  775 / 0.37 / 0.29 &  1585 / 0.24 / 0.19 &  631 / 0.28 / 0.23 &  740 / 1.01 / 0.30 &  7464 / 0.31 / 0.34 \\
\bottomrule
    \toprule
    \multicolumn{9}{c}{Predicted Peak Performance} \\
    \toprule
  \textbf{RGZ}  &  \textbf{Prediction} &          1\_1 &          1\_2 &         1\_3 &          2\_2 &         2\_3 &         3\_3 &      Summary \\
  \textbf{Radio CL} & \textbf{Result} &              &              &             &              &             &             &              \\
  \midrule
\textbf{0.6-0.7} & \textbf{Incorrect} &    69 / 0.88 / 0.25 &     8 / 0.64 / 0.08 &   35 / 0.77 / 0.16 &    74 / 0.60 / 0.10 &  110 / 0.66 / 0.16 &  204 / 0.71 / 0.27 &   500 / 0.71 / 0.23 \\
        & \textbf{Correct} &    65 / 0.10 / 0.14 &    93 / 0.16 / 0.13 &   15 / 0.33 / 0.12 &   326 / 0.32 / 0.12 &   68 / 0.37 / 0.10 &   75 / 0.41 / 0.07 &   642 / 0.29 / 0.15 \\
\textbf{0.7-0.8} & \textbf{Incorrect} &    27 / 0.85 / 0.23 &     4 / 0.62 / 0.13 &   23 / 0.76 / 0.31 &    82 / 0.59 / 0.10 &  126 / 0.64 / 0.15 &  200 / 0.66 / 0.20 &   462 / 0.66 / 0.19 \\
        & \textbf{Correct} &    39 / 0.10 / 0.13 &    57 / 0.17 / 0.12 &   19 / 0.37 / 0.12 &   412 / 0.30 / 0.12 &   95 / 0.36 / 0.10 &   92 / 0.40 / 0.10 &   714 / 0.30 / 0.14 \\
\textbf{0.8-0.9} & \textbf{Incorrect} &    20 / 0.84 / 0.23 &     3 / 0.76 / 0.08 &   23 / 0.78 / 0.29 &    77 / 0.60 / 0.07 &   98 / 0.66 / 0.24 &  106 / 0.68 / 0.24 &   327 / 0.68 / 0.22 \\
        & \textbf{Correct} &    41 / 0.11 / 0.13 &    75 / 0.15 / 0.13 &   19 / 0.35 / 0.12 &   435 / 0.30 / 0.11 &   77 / 0.37 / 0.09 &   41 / 0.39 / 0.11 &   688 / 0.29 / 0.14 \\
\textbf{0.9-1.0} & \textbf{Incorrect} &   386 / 0.87 / 0.24 &   146 / 0.64 / 0.12 &  441 / 0.74 / 0.22 &    20 / 0.57 / 0.09 &   33 / 0.65 / 0.11 &   15 / 0.66 / 0.19 &  1041 / 0.77 / 0.23 \\
        & \textbf{Correct} &  1300 / 0.09 / 0.12 &  1400 / 0.17 / 0.13 &  200 / 0.34 / 0.11 &   159 / 0.30 / 0.12 &   24 / 0.37 / 0.10 &    7 / 0.41 / 0.05 &  3090 / 0.16 / 0.15 \\
\bottomrule
\textbf{Summary} & {} &  1947 / 0.29 / 0.38 &  1786 / 0.21 / 0.19 &  775 / 0.61 / 0.27 &  1585 / 0.35 / 0.16 &  631 / 0.53 / 0.21 &  740 / 0.60 / 0.24 &  7464 / 0.37 / 0.30 \\
\bottomrule
    \end{tabular}
    \captionof{table}{An overview of the predictive performance of the two \texttt{RandomForestRegressor}s predicting the number of components (top) and number of peaks (bottom) which have been trained using the preprocessed images that were provided to PINK. The results have been broken down by the RGZ Radio CL, the RGZ label and whether the prediction result is correct if a hard classification scheme is used. For each intersection (including the summary column and row) we include the number of \revisiontwo{objects} in that subset, their average absolute distance from zero (a perfect prediction) and the standard deviation of the absolute distance from zero. \label{tab:rgz_rfr_pink_predictor}}
  \end{figure*}

In Figure~\ref{fig:mse_hist_pink} we show the difference between the predicted number of components and peaks and the corresponding RGZ label. Inspecting the difference between the predicted and labelled number of peaks, there is a clear positive bias. Although this was also seen in Figure~\ref{fig:mse_hist}, the effect seen is much more pronounced and extends to lower ranked consensus levels. Unlike Figure~\ref{fig:mse_hist} however, examining the difference between the predicted and labeled number of peaks shows strong features. Not only is there a strong peak set near the zero made up of the high consensus level objects with a positive skew, there are additional features approximately centered at --0.5 and 0.5. 

The presences of the features in Figure~\ref{fig:mse_hist_pink} indicates that the \texttt{RandomForestRegressor} is not capable of handling the affine transforms applied to resolved \revisiontwo{objects} with complex morphologies. Performing similar experiments with different \texttt{RandomForestRegressor} options set produced consistent behaviour. We can therefore consider PINK as a useful tool to assist with not only reducing the feature space by a factor of $\sim248$, but also with producing more reliable classification or knowledge transfer tools by factoring out affine transforms of similar morphologies. 

An important point to emphasize is that if the fuzzy labels of Figure~\ref{fig:mse_hist_pink} were rounded to their nearest label, then 78\% and 68\% of the data-set would be centered at zero. At a glance, these are comparable to the results of applying the \texttt{RandomForestRegressor} to the PINK produced likelihood matrices, and in fact the total fraction of \revisiontwo{objects} with the correct number of components is higher than the 74\% seen in Figure~\ref{fig:mse_hist}. However, when the fuzzy labels are used to calculate the sum of the absolute distance of each \revisiontwo{object}, the residual distance for the predicted components and peaks are 915.71 and 1059.06, respectively. Both of these are at least $\sim$15\% higher than the distances measured using the PINK based regressor method, with the number of components being 30\% higher. The use of a hard classification scheme concealed the residual error of the \revisiontwo{object's} predicted labels. It may potentially be more useful to consider a soft classification scheme to better capture anomalous behavior in labels and their distribution. 

\revisiontwo{We include in Table~\ref{tab:rgz_rfr_pink_predictor} an overview of component and peak \texttt{RanddomForestRegressor} predictors, broken into RGZ label and radio CL subsets.} Although there are some classes that perform similarly to models trained on PINK produced products when a hard classification scheme is adopted (Tables~\ref{tab:rgz_rfr_predictor} and \ref{tab:rgz_rfr_cl_predictor}), we highlight that the average absolute distance is higher for almost all items. This measure indicates that there was a large amount of disagreement among the ensemble of trees forming the random forest.   

\subsection{Learnt Multi-Wavelength Features}

Identifying and associating resolved discrete components of a single intrinsic object is a difficult problem, and one that is further complicated when attempting to incorporate multi-wavelength data. 

In the radio domain, \citet{2017PASA...34....3L} attempt to use spectral index information to associate related radio components between different catalogues. Across a small radio frequency range, the spectral energy distribution of radio sources is characterized well by a power law, which is incorporated as an additional metric when assessing whether multiple discrete components at a particular frequency are related to a single component at another frequency. 

Expanding this cross-cataloguing to a wider wavelength space, further complicates the problem. Different physical processes often mean that components of complex sources are separated and appear to be distinct. For instance, radio lobes of AGN can be separated by some distance from the host galaxy, which is often an unresolved point source at infra-red wavelengths.

Human pattern recognition, being well suited for the problem, has already been employed as a solution. However, even harnessing the collective power of citizen scientists through RGZ and related projects, it is unlikely to scale to the millions of objects to be discovered by EMU and the Square Kilometre Array (SKA) era of radio astronomy. 

Alternative codes show promise at this specific multi-wavelength cross-cataloging problem. The Likelihood Ratio in PYthon \citep[LRPY; ][]{2018MNRAS.473.4523W} package implements a modified version of the likelihood method \citep{1975AN....296...65R} to attempt to solve the problem for single radio sources, but cannot yet tackle multi-component radio sources. 

\citet{2015MNRAS.451.1299F} approach the problem of multi-component sources within a Bayesian framework. Constructing a model of AGN and their morphologies, they are able to cross-catalogue components of radio objects, such as cores or lobes, to IR components which are not necessarily co-located in the same region of sky. This method incorporates a `prior distribution' when assessing the potential match. The prior distribution incorporates known information about a particular model or morphology. 

\begin{figure}
  \includegraphics[width=\linewidth, clip=true, trim=0.cm 0.cm 0.cm 0cm]{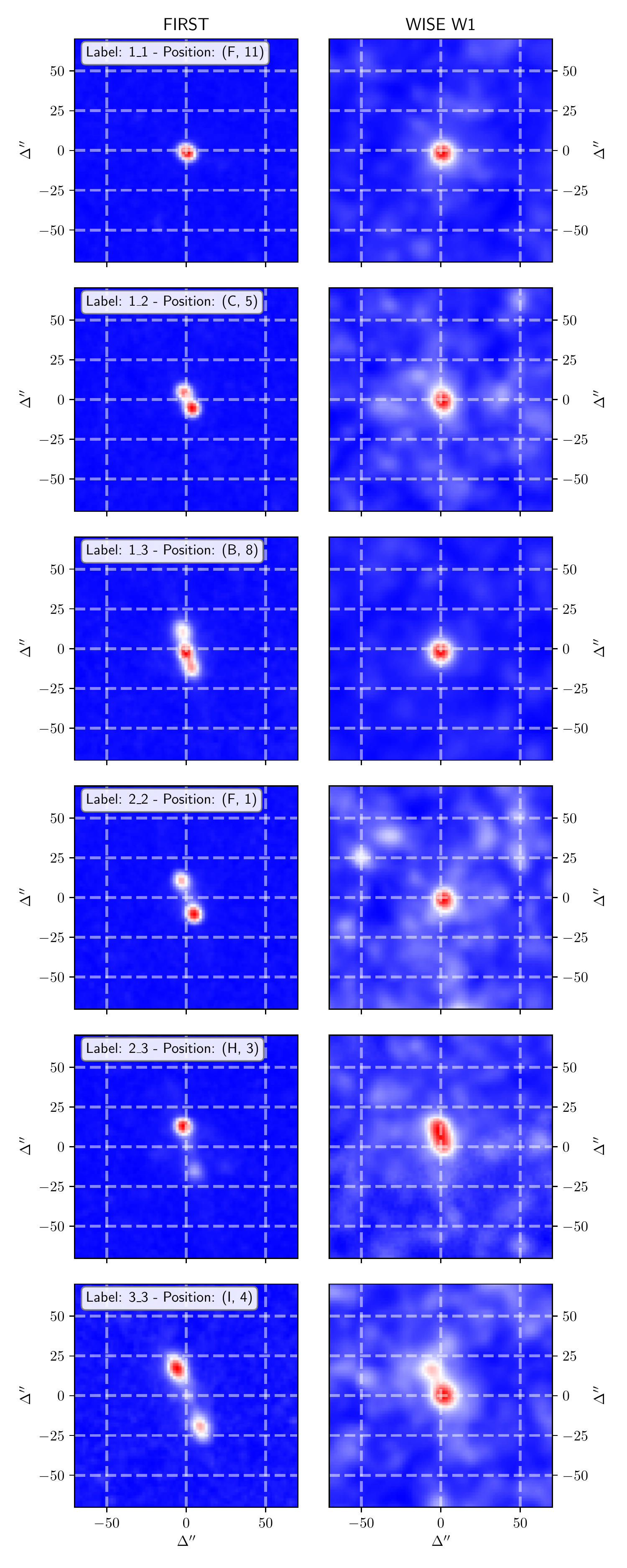}
  \caption{An example of trained neurons produced by PINK showing the learnt features for the position of each of the red markers in Figure~\ref{fig:median_prob}. The left hand column shows the radio data and the right hand column shows the infrared data. In these examples, the infrared sources lies between two radio lobes. The overlaid annotation in the upper region of each panel in the left column indicates the corresponding label and the position of the neuron in Figure~\ref{fig:WISE_Norm}. Each panel has had a square root transform applied to better emphasis their learnt features. \label{fig:example_neurons}}
\end{figure}

Using a PINK trained SOM based on multi-wavelength image cubes of objects, it should be possible to craft additional prior distribution information. To demonstrate this, we extracted six neurons from the SOM shown in Figure~\ref{fig:WISE_Norm}, where each neuron corresponding to a peak pixel of the median likelihood matrices presented in Figure~\ref{fig:median_prob}. These six neurons are presented in Figure~\ref{fig:example_neurons}. We emphasize that the features in each panel are constructed in an unsupervised manner by PINK, and can be considered ideal prototypes of a set of \revisiontwo{objects} in the input data set. PINK has identified not only a range of radio features, including AGN produced radio lobes, but also where the corresponding IR host is located in the WISE channel. 
\revision{These IR components were constructed from images which had no sigma noise clipping or masking of features unrelated to the centered component. }

\begin{figure}
  \centering
  \includegraphics[width=\linewidth, clip=true, trim=0.cm 0.cm 0.cm 0cm]{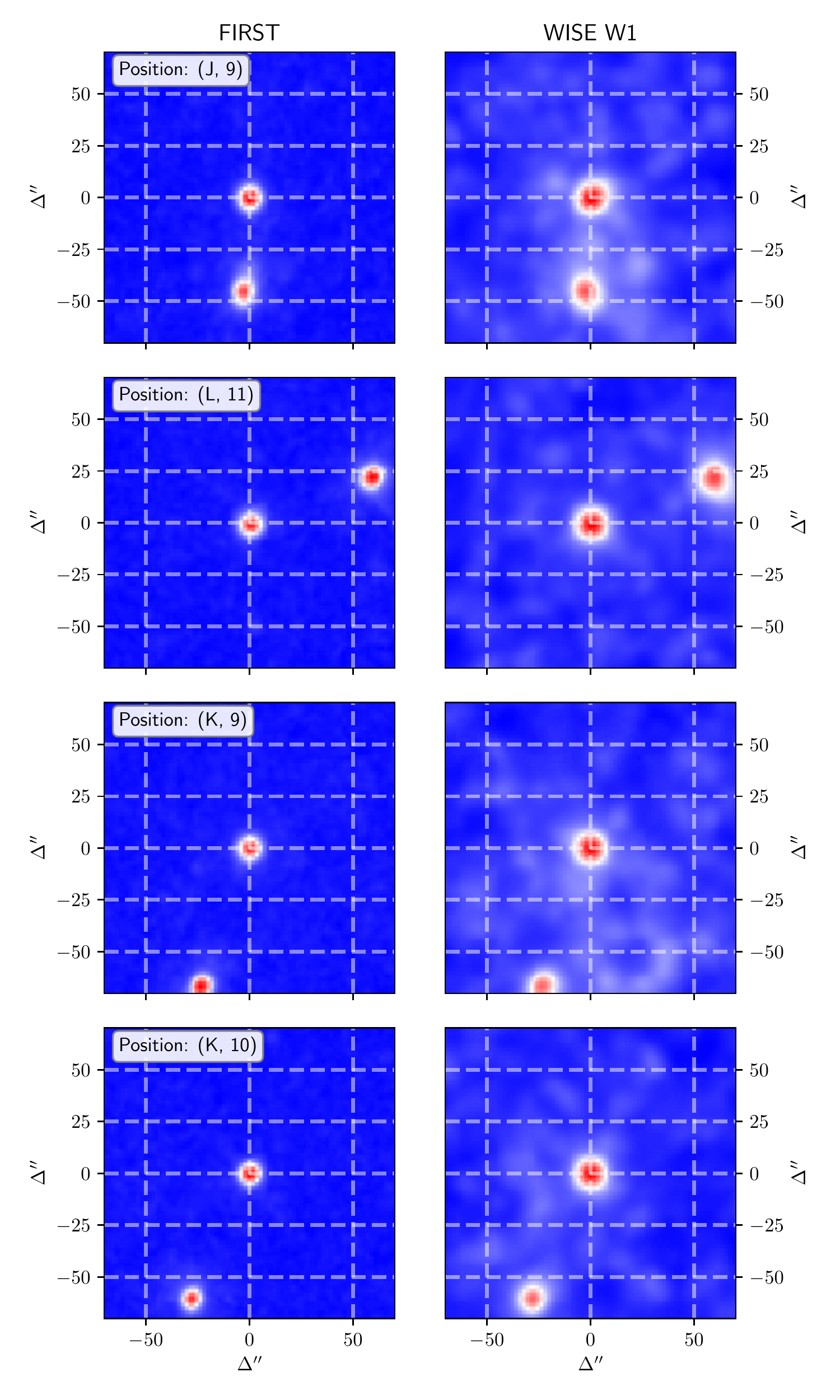}
  \caption{Examples of selected neurons from Figure~\ref{fig:WISE_Norm} with what could be classified as independent galaxies. The left hand column shows the radio data and the right hand column shows the infrared data. In these examples, an unrelated source in the field is detected in both the radio and infrared  images.\label{fig:discrete_objects} }
\end{figure}

There are also a set of neurons which have learnt to recognize \revisiontwo{objects} which could have a unrelated, nearby galaxies within the field. These neurons, which we include as Figure~\ref{fig:discrete_objects}, show two separate components in both the FIRST and WISE channels. This could suggest that these objects are intrinsically unrelated to one another, as they both have a clear IR component but just happen to be within a close proximity to one another by chance.  

We propose that if the training sample of objects and their morphologies are an accurate representative samples of data to be cross-catalogued, then the trained neurons themselves can be used as archetype morphologies. Classifying the features across the different multi-wavelength channels of each neuron would be fairly trivial \revision{when compared to the task of classifying the raw input source images directly}, even for a large SOM \revisiontwo{lattice}. The weighting updates throughout the training stage of PINK minimises noise. With automated source finders, this could be performed with little intervention from an expert. Alongside the complete set of \revision{likelihood} matrices of \revisiontwo{objects} and the extracted set of features across all neurons, a robust set of well characterized prior functions can be incorporated into Bayesian based cross-cataloging methods. The smooth nature of the SOMs would allow for interpolation to be implemented in constructed prior functions - an important characteristic for distance separating candidate features. 

Being an unsupervised method, these priors can be updated with little human effort for different combinations of surveys. Surveys of any wavelength have their own set of sensitivities, limitations and biases. Training any ML method on one data-set and applying its learnt knowledge onto another can easily introduce potential issues, reducing its predictive power and ultimate use. This could be particularly troublesome for radio-continuum data, where the telescope array configuration and observing frequency not only influences the type of angular scales and physical processes the instrument is sensitivity to, but also affects the point spread function that has to be modelled and removed from the imaging process. Understanding these effects would be critical to ensure the transfer of knowledge between different data-sets using previously trained ML methods.


Developing and extracting useful multi-wavelength features in an unsupervised manner to craft prior distributions for cross-cataloging problems offers an ideal characteristic, where the only cost to retrain is computation time. 

\revision{An alternate approach to this cross-cataloguing problem is to exploit the position of the IR host in each of the neurons. Assume a SOM has been trained based on image cubes (with radio and IR information) for positions centered on islands of pixels found by a source finder on a radio image. Once trained, neurons will show locations of radio components and the IR host. Knowing the radio position of an \revisiontwo{object} and the offset of the IR host in the BMU neuron with respect to this position, then the \revisiontwo{corresponding sky} position of a IR host for each radio component can be estimated. An internal match looking for common estimated IR host locations could be used to identify related object, linking together their radio and IR host components. This method is being investigated by Hopkins et al. (in prep.). }

\section{Conclusion and Future Work}
\label{sec:conclusion}

We have used the PINK software to produce a multi-channel SOM using 100,000 source images based on the RGZ DR1. We find that:

\begin{enumerate}
  \item PINK is able to produce a SOM that exhibits a range of morphologies that are representative prototypes of the training \revisiontwo{objects} used,
  \item Labels produced by RGZ participants are able to be clustered on the surface of the constructed SOM, 
  \item Similarity measures produce by PINK can be used as a basis of \revisiontwo{object} classification and can improve knowledge transfer, 
  \item A \texttt{RandomForestRegressor} was used as a mechanism to efficiently derive and apply a soft classification scheme based on labeled training \revisiontwo{objects},
  \item Adding additional features alongside the PINK produced similarity measures improved the predictive power of the \texttt{RandomForestRegressor},   
  \item A soft classification scheme that avoids discrete intervals or boundaries should be considered in the near future as it can reveal hidden biases or inconsistencies, 
  \item The similarity measure produced by PINK assisted in improving the classification performance of labeled training data when the absolute distance was considered as a metric, and
  \item Physically meaningful features across multiple wavelengths can be constructed by PINK and potentially used to craft prior distribution functions for Bayesian based methods. 
\end{enumerate}

At the time of writing, PINK does not make available to the user the transform matrix that is used to  place an input image onto the SOM \revisiontwo{lattice}, although this feature is now planned to be included in a future release. RGZ maintains information of the position of all user clicks as the citizen scientists are classifying features within an \revisiontwo{object's} image. An interesting application of these transforms would be applying the transform information derived by PINK onto user click information maintained by RGZ to project them \revisiontwo{onto a set of trained prototypes}. Each prototype should then show robustly characterized regions highlighting important features within each neuron in a manner similar to a kernel density estimator. Although an individual \revisiontwo{object} image may only have been inspected tens of times by RGZ scientists through the Zooniverse, when transformed and placed onto a trained SOM surface there is potentially information on thousands of user clicks,  highlighting consistent features within each neuron. We plan to investigate this once our feature request has been added to a future version of PINK.  

\revision{We are also investigating whether hierarchically structured SOMs, similar to \citet{dittenbach2000growing}, can be used to segment data-sets into different classes in a unsupervised (or semi-supervised) manner which can then be used as the basis of more reliable classifications scheme. Our \texttt{RandomForestRegressor} performed poorly on \revisiontwo{objects} with more complex RGZ labels. We suspect that this is a combination of effects, including inconsistent responses among the citizen scientists (which is characterised by the CL) and complex morphologies being to compressed on the SOM surface. A hierarchical SOM may produce a better set of features for other ML methods, including random forest classifiers, to operate on.  }

\section*{Acknowledgements}
\revision{We thank the anonymous referees whose thoughtful feedback improved the presentation and quality of this manuscript. }

KP and EH gratefully acknowledge the support of the Klaus Tschira Foundation. 

Partial support for LR comes from U.S. National Science Foundation grant AST17-14205 to the University of Minnesota.

This  publication  has  been  made  possible  by  the  participation  of  more  than
250,000 volunteers in the Galaxy Zoo Project. The data in this paper are the 
result of the efforts of the  Radio  Galaxy  Zoo volunteers,  without  whom  none  of  this  work  would  be  possible. Their efforts are individually acknowledged at \url{http://rgzauthors.galaxyzoo.org}. 

The authors would like to thank Chen Wu for providing a publicly accessible repository for code and initial training data and for assistance throughout the project.

This publication makes use of data products from the {\it Wide-field Infrared Survey Explorer} and the {\it Spitzer Space Telescope}.  The {\it Wide-field Infrared Survey Explorer} is a joint project of the University of California, Los Angeles, and the Jet Propulsion Laboratory/California Institute of Technology, funded by the National Aeronautics and Space Administration.  SWIRE is supported by NASA through the SIRTF Legacy Program under contract 1407 with the Jet Propulsion Laboratory.  This publication makes use of radio data from the Australia Telescope Compact Array and the Karl G. Jansky Very Large Array (operated by NRAO).   The Australia Telescope Compact Array is part of the Australia Telescope, which is funded by the Commonwealth of Australia for operation as a National Facility managed by CSIRO.   The National Radio Astronomy Observatory is a facility of the National Science Foundation operated under cooperative agreement by Associated Universities, Inc.

Code used throughout the data preparation, PINK operation stages and production of Figures are hosted at \url{https://github.com/tjgalvin/Pink_Experiments}. 

The \texttt{astropy} \citep{2013A&A...558A..33A,2018arXiv180102634T} affiliated \texttt{python} based \texttt{reproject}\footnote{\url{https://reproject.readthedocs.io/en/stable/}} module was used for pre-processing of the data.

\bibliographystyle{aasjournal}
\bibliography{som_rgz}

\end{document}